\documentclass[sn-mathphys-num]{sn-jnl}

\usepackage{graphicx}%
\usepackage{multirow}%
\usepackage{amsmath,amssymb,amsfonts}%
\usepackage{amsthm}%
\usepackage{mathrsfs}%
\usepackage[title]{appendix}%
\usepackage{xcolor}%
\usepackage{textcomp}%
\usepackage{manyfoot}%
\usepackage{booktabs}%
\usepackage{algorithm}%
\usepackage{algorithmicx}%
\usepackage{algpseudocode}%
\usepackage{listings}%
\usepackage{multirow}
\usepackage{tabu}
\usepackage{bm}
\usepackage{float}
\usepackage{array}
\newcolumntype{P}[1]{>{\centering\arraybackslash}p{#1}}
\usepackage{glossaries} 

\makeglossaries%
\setkeys{glslink}{hyper=false}

\definecolor{myred}{RGB}{0,0,0}

\newacronym{SNR}{SNR}{Signal to Noise Ratio}
\newacronym{DNN}{DNN}{Deep Neural Network}
\newacronym{LPC}{LPC}{Linear Predictive Coding}
\newacronym{MFCC}{MFCC}{Mel-Frequency Cepstral Coefficients}
\newacronym{FIR}{FIR}{Finite Impulse Response}
\newacronym{VAD}{VAD}{Voice Activity Detection}
\newacronym{SSL}{SSL}{Sound Source Localisation}
\newacronym{SER}{SER}{Speech Emotion Recognition}
\newacronym{ANN}{ANN}{Artificial Neural Network}
\newacronym{IPD}{IPD}{Interaural Phase Difference}
\newacronym{ILD}{ILD}{Interaural Level Difference}
\newacronym{HRTF}{HRTF}{Head-related Transfer Function}
\newacronym{HRIR}{HRIR}{Head-related impulse responses}
\newacronym{HCI}{HCI}{Human Computer Interaction}
\newacronym{i.i.d}{i.i.d}{independent and identically distributed}
\newacronym{HMM}{HMM}{Hidden Markov Model}
\newacronym{DCT}{DCT}{Discrete Cosine Transform}
\newacronym{GMM}{GMM}{Gaussian Mixture Model}
\newacronym{LEM}{LEM}{Laplacian Eigenmaps}
\newacronym{VR}{VR}{Virtual Reality}
\newacronym{CNN}{CNN}{Convolutional Neural Network}
\newacronym{RNN}{RNN}{Recurrent Neural Network}
\newacronym{GRU}{GRU}{Gated Recurrent Unit}
\newacronym{LSTM}{LSTM}{Long Short-term Memory}
\newacronym{GeMAPS}{GeMAPS}{Geneva Minimalistic Acoustic Parameter Set}
\newacronym{CCC}{CCC}{Concordance Correlation Coefficient}
\newacronym{IEMOCAP}{IEMOCAP}{Interactive Emotional Dyadic Motion Capture}
\newacronym{RECOLA}{RECOLA}{REmote COLlaborative and Affective}
\newacronym{GPU}{GPU}{Graphical Processing Unit}
\newacronym{AVEC}{AVEC}{Audio/Visual Emotion Challenge and Workshop}
\newacronym{ECG}{ECG}{Electrocardiogram}
\newacronym{EDA}{EDA}{Electro-dermal activity}
\newacronym{TTS}{TTS}{Text-to-speech}
\newacronym{TDNN}{TDNN}{Time-Delay Neural Network}
\newacronym{NLP}{NLP}{Natural Language Processing}
\newacronym{LOSO}{LOSO}{Leave-one-speaker-out}
\newacronym{LOGO}{LOGO}{Leave-one-group-out}
\newacronym{MSE}{MSE}{Mean Squared Error}
\newacronym{MAE}{MAE}{Mean Absolute Error}
\newacronym{PCC}{PCC}{Pearson Correlation Coefficient}
\newacronym{ReLU}{ReLU}{Rectified Linear Unit}
\newacronym{DiCCOSER-CS}{DiCCOSER-CS}{Dilated Causal Convolution Only Speech Emotion Recognition with Context-Stacking}
\newacronym{WA}{WA}{Weighted Accuracy}
\newacronym{UA}{UA}{Unweighted Accuracy}
\newacronym{PCA}{PCA}{Principal Component Analysis}
\newacronym{STFT}{STFT}{Short-time Fourier transform}
\newacronym{ASR}{ASR}{Automatic Speech Recognition}
\newacronym{CNN-LSTM}{CNN-LSTM}{CNN-LSTM}
\newacronym{WCCN}{WCCN}{Within-Class Covariance Normalisation}
\newacronym{Deep-WCCN}{Deep-WCCN}{Deep-Within-Class Covariance Normalisation}
\newacronym{RAVDESS}{RAVDESS}{Ryerson Audio-Visual Database of Emotional Speech and Song}
\newacronym{ESD}{ESD}{Emotional Speech Dataset}
\newacronym{Emo-DB}{Emo-DB}{Berlin Database of Emotional Speech}
\newacronym{eGeMAPS}{eGeMAPS}{Extended Geneva Minimalistic Acoustic Parameter Set}
\newacronym{SGD}{SGD}{Stochastic Gradient Descent}
\newacronym{RF}{RF}{Random Forest}
\newacronym{NN}{NN}{Neural Network}
\newacronym{ITD}{ITD}{Interaural Time Difference}
\newacronym{RTF}{RTF}{Relative Transfer Function}
\newacronym{LPP}{LPP}{Locality-preserving projection}
\newacronym{LBO}{LBO}{Laplace-Beltrami Operator}
\newacronym{MDS}{MDS}{Multidimensional scaling}
\newacronym{TF}{TF}{time-frequency}
\newacronym{RIR}{RIR}{Room Impulse Response}
\newacronym{SCE}{SCE}{Supervised Contrastive Embedding}
\newacronym{WSCE}{WSCE}{Weakly supervised Contrastive Embedding}
\newacronym{MLE}{MLE}{Maximum likelihood estimation}
\newacronym{GBDT}{GBDT}{Gradient Boosted Decision Trees}
\newacronym{ADC}{ADC}{Analog-to-digital converter}
\newacronym{SVM}{SVM}{Support Vector Machine}
\newacronym{i.e.}{i.e.}{\emph{id est}, that is}
\newacronym{e.g.}{e.g.}{\emph{exempli gratia}, for example}
\newacronym{DFT}{DFT}{Discrete Fourier transform}
\newacronym{t-SNE}{t-SNE}{t-distributed Stochastic Neighbor Embedding}
\newacronym{GAN}{GAN}{Generative Adversarial Network}

\theoremstyle{thmstyleone}%
\theoremstyle{thmstyletwo}%

\theoremstyle{thmstylethree}%

\begin{document}
	
	\title[Article Title]{End-to-end transfer learning for speaker-independent cross-language and cross-corpus speech emotion recognition}
	
	
	\author*[1]{\fnm{Duowei} \sur{Tang}}\email{duowei.tang@kuleuven.be}
	
	\author[2]{\fnm{Peter} \sur{Kuppens}}\email{peter.kuppens@kuleuven.be}
	
	\author[3]{\fnm{Lucca} \sur{Geurts}}\email{lucca.geurts@kuleuven.be}
	
	\author[1]{\fnm{Toon} \sur{van Waterschoot}}\email{toon.vanwaterschoot@esat.kuleuven.be}
	
	\affil*[1]{\orgdiv{Department of Electrical Engineering (ESAT-STADIUS/ETC)}, \orgname{KU Leuven}, \orgaddress{\street{Kasteelpark Arenberg 10}, \city{Leuven}, \postcode{3001}, \country{Belgium}}} 
	
	\affil[2]{\orgdiv{Faculty of Psychology and Educational Sciences}, \orgname{KU Leuven}, \orgaddress{\street{Tiensestraat 102}, \city{Leuven}, \postcode{3000}, \country{Belgium}}}
	
	\affil[3]{\orgdiv{Department of Computer Science (Group T Leuven Campus)}, \orgname{KU Leuven}, \orgaddress{\street{Andreas Vesaliusstraat 13}, \city{Leuven}, \postcode{3000}, \country{Belgium}}}
	
	
	
	\abstract{Data-driven models achieve successful results in \gls{SER}. However, these models, which are often based on general acoustic features or end-to-end approaches, show poor performance when the testing set has a different language than the training set (i.e. in a cross-language setting) or when these sets are taken from different datasets (i.e. in a cross-corpus setting). To alleviate these problems, this paper presents an end-to-end \gls{DNN} model based on transfer learning for cross-language and cross-corpus \gls{SER}. We use the wav2vec 2.0 pre-trained model to transform audio time-domain waveforms from different languages, different speakers and different recording conditions into a feature space shared by multiple languages, thereby reducing the language variabilities in the speech embeddings. Next, we propose a new \gls{Deep-WCCN} layer that can be inserted into the \gls{DNN} model and aims to reduce other variabilities including speaker variability, channel variability and so on. The entire model is fine-tuned in an end-to-end manner on a combined loss and is validated on datasets from three languages (i.e. English, German, Chinese). Experimental results show that our proposed method not only outperforms the baseline model that is based on common acoustic feature sets for \gls{SER} in the within-language setting, but also significantly outperforms the baseline model for the cross-language setting. In addition, we also experimentally validate the effectiveness of \gls{Deep-WCCN}, which can further improve the model performance. Next, we show that the proposed transfer learning method has good data efficiency when merging target language data into the fine-tuning process. The model speaker-independent \gls{SER} performance increases with up to 15.6\% when only 160\,s of target language data is used. Finally, when comparing with the results in recent literatures which use the same testing datasets, our proposed model shows significantly better performance than other state-of-the-art models in cross-language \gls{SER}.}

	
	
	
	\maketitle
	
	\section{Introduction}
	\label{sec:xlser_intro}
	
	The emotions we daily experience determine for a large part our mental flourishing and suffering. Happiness, or psychological well-being, relies greatly on how people experience positive and negative emotions in their lives \cite{Krueger2014}. Modern \gls{HCI} systems use image/video, speech, and physiological signals to determine people's emotion \cite{Saxena2020}. Vocal expression is a direct and affectionate way of expressing emotions that has the advantage of being more accessible than image/video and physiological signals, for which a careful camera positioning or a well-worn wearable device is needed. As a result, \gls{SER} is widely used in many applications, such as, an in-car board system that can provide aids or resolve errors in the communication according to the driver's emotion \cite{Schuller2004}, a diagnostic tool that uses the user's speech emotion to provide diagnostic information to the physiotherapist \cite{France2000}, and an assistant robot that can provide emotional communication \cite{Castillo2018}. 
	
	\gls{SER} using data-driven models has been successful in recognising emotions~\cite{Schuller2009,ElAyadi2011,Xiao2017,Zhao2019a,Zhang2019,Tzirakis2018,Tang2021}. However, many data-driven models rely strongly on the mechanism underlying the generation of the data (i.e. the \gls{i.i.d} assumption), and may hence fail when the testing data is taken from a different distribution than the training data. Alike other speech-related tasks, \gls{SER} becomes highly challenging in cross-language, cross-corpus, and cross-speaker scenarios \cite{Bhaykar2013, Hozjan2003}. To facilitate these challenges, traditional \gls{SER} approaches use low-level-descriptor features that have been selected and grouped in the \gls{GeMAPS} feature set, including various acoustic features such as frequency-related
	(e.g. pitch, formant), energy-related (e.g. loudness) and spectral (e.g. spectral slope) features. In \cite{Xiao2017}, the authors use low-level-descriptor features in cross-language \gls{SER}, showing that people speaking different languages may express emotion in a similar way at the low-level signal level. To extract features that contain higher-level information, the i-vector proposed in \cite{Dehak2011} for speaker verification, has been further extended to \gls{SER} where the ``emotion i-vector'' shows robustness in an English-German cross-language \gls{SER} setting~\cite{Desplanques2018}.
	
	In recent work, the complicated feature design process is taken over by a unified \gls{DNN} model that enables end-to-end learning from raw data or shallow features \cite{Zhao2019a,Zhang2019,Tzirakis2018,Tang2021}. Robust features are then learned from data to minimise an overall loss towards emotion recognition. Many works achieve good \gls{SER} performance on a single dataset \cite{Zhao2019a,Zhang2019,Tzirakis2018,Tang2021}, but the model performance degrades dramatically in cross-language and cross-corpus scenarios \cite{Hozjan2003}. To alleviate this problem, several transfer learning techniques have been applied. In \cite{Neumann2018,Latif2018}, the authors pre-train a neural network with emotional datasets from one or two languages, then fine-tune the model with a small portion of the target language data, which shows a performance improvement compared to when the model is trained with one language and tested on another languages (i.e. in the cross-language setting). In \cite{Gerczuk2021}, a similar idea to pre-train a \gls{DNN} model using multiple emotional datasets is proposed, and the authors propose to fine-tune only a task-specific parallel residual adapter which achieves increased \gls{SER} performance on each target task while keeping the amount of parameters to be updated low. 
	
	Other than to adapt the pre-trained model to the target language, one can also learn language-invariant or corpus-invariant features. This can be done by transforming the input space to a common subspace among each corpus. In \cite{Song2019}, Song firstly proposed to transform the source and target input space to a common  subspace where the input source-target neighbourhood relations are preserved. Zhang and Song later updated this framework to include sparsity and add discriminative power to the learning of the subspace \cite{Zhang2020}. To enable non-linear modelling with the \gls{DNN}, the transformation to the common subspace is merged into an encoder neural network with a discriminator, and the learning is adversarial between the encoder and the discriminator where the encoder transforms source and target inputs to embeddings that will fool the discriminator, and then the discriminator needs to find the true related inputs (i.e. from the source dataset or the target dataset) \cite{Latif2019,Liang2019}. However, these methods require the definition of the ``source'' and ``target'' language beforehand, and their application is limited to those two languages. In \cite{Gideon2019}, instead of using a binary discriminator, a Wasserstein distance \cite{Arjovsky2017} is used to measure the distances between target and source embeddings, and the method minimises the distances between them. 
	
	\gls{DNN} models are data-driven, incorporating large amounts and a wide variety of training data generally enhances their generalisation capability. Methods that leverage large volumes of unlabelled or synthesised data are emerging. Pepino et. al. employ a large-scale self-supervised pre-trained speech representation model, wav2vec 2.0 \cite{Baevski2020}, to extract general speech features. These features are used for within-corpus \gls{SER}, showing significant performance improvements over traditional features. However, cross-language and cross-corpus performance are not evaluated in this work \cite{Pepino2021}. In \cite{sharma2022multi}, the author utilises a variety of emotional datasets from multiple languages to fine-tune the wav2vec 2.0 model within a multi-task learning framework. The use of multilingual emotional data improves \gls{SER} performance across languages included in training, though cross-language and cross-corpus results are still missing. In \cite{ma2024emotion2vec}, Ma et. al. adopt a novel self-supervised learning method based on online distillation \cite{grill2020bootstrap} and incorporate a wide range of emotional datasets. Their method learns robust and generalised emotion representations that can be applied across different emotion recognition tasks (e.g., song emotion recognition, emotion prediction in conversation, and sentiment analysis) and languages. To synthesise data for training \gls{SER} models, the authors in \cite{latif2022self} propose a \gls{GAN} based framework to generate synthetic emotional data for fine-tuning the \gls{SER} model.
	
	We propose an end-to-end transfer learning framework to facilitate cross-language \gls{SER}. This method lies in-between the transfer learning method and the common subspace method. First, the proposed model uses a wav2vec 2.0 feature extractor which has been trained in a self-supervised manner on about 56,000 hours of raw speech waveforms from 53 languages. Since the pre-trained wav2vec 2.0 feature extractor learns contextual structures across various speech utterances, it may capture common speech factors among different languages, hence transforming the speech waveform inputs to a common speech subspace which is shared across languages \cite{Conneau2020}, and in which the corresponding speech embeddings are obtained. Next, a statistical pooling layer is applied that pools a sequence of the embeddings into one feature vector per utterance. The utterance-level feature vectors are then reduced in dimensionality, and a \gls{Deep-WCCN} operation is applied to compensate for other variabilities (i.e. other than language variabilities). Finally, the compensated feature is used to predict emotion class with a simple linear classifier (constructed by a dense neural network layer). The model is fine-tuned on three emotional speech datasets with English, German, and Chinese languages using a summation of the supervised cross-entropy loss and the original wav2vec 2.0 loss. We evaluate its leave-one-language-out performance on unseen speakers, and the experimental results show that the proposed framework largely increases the cross-language \gls{SER} performance compared to two well-known feature sets extracted using openSMILE for \gls{SER}. It also largely outperforms many recent approaches to both within-language and cross-language \gls{SER}. Finally, additional experiments on the effectiveness of the \gls{Deep-WCCN} operation, and the effects of merging small amounts of target language speech data into the training set will be presented.
	
	The paper is structured as follows. First, Section\,\ref{sec:xlser_relatedworks} provides a brief overview about the most recent studies related to our work. Then we introduce the proposed model structure in Section\,\ref{sec:xlser_method}, and propose some modifications to a similar approach in \gls{Deep-WCCN} \cite{Eghbal-zadeh2018}. In Section\,\ref{sec:xlser_exp}, we describe the experiment datasets and the experimental settings. After that, we will present the simulation results and discuss these in Section\,\ref{sec:xlser_results}. Finally, Section\,\ref{sec:xlser_conclusions} presents the conclusions and suggestions for future work.
	
	\section{Related works}
	\label{sec:xlser_relatedworks}
	\subsection{\gls{SER} features}
	Traditional speech information retrieval systems use hand-crafted features such as the \gls{MFCC} features and the \gls{LPC} features \cite{Hafen}, and those features are designed to largely reduce the data dimensionality and to encode temporal and spectral structures per time frame in a speech recording. A group of low-level descriptor features such as pitch, formant frequencies and bandwidths, loudness, and spectral slope have been selected and compared in \cite{Schuller2009} for \gls{SER}. Reynolds and Rose later proposed to use \glspl{GMM} to model the speakers' \gls{MFCC} distribution \cite{Reynolds1995}. Each individual Gaussian component of a \gls{GMM} encodes a general acoustic class (e.g., vowels, nasals, or fricatives), and the spectral shape of the acoustic class can be characterised by the \gls{GMM} component mean vector and the \gls{GMM} component covariance matrix. The concatenation of \gls{GMM} component mean vectors is a representation that contains abstract information such as speaker identity. This high-dimensional representation (referred to as the supervector in \cite{Dehak2011}) is then used to extract the i-vector which encodes the speaker and channel information in a total variability subspace \cite{Dehak2011}. However, these hand-crafted features are not designed for \gls{SER} specifically, and may contain dominant misleading factors (e.g., speaker identities, language factors) which leads to poor \gls{SER} performance. This problem is tackled in \cite{Desplanques2018} where a two-step approach is proposed. The approach first compensates for the speaker variability in the supvervector space, then an ``emotion i-vector'' that encodes the variabilities of each emotion supervector centred around a maximum likelihood emotion supervector is extracted. 
	
	\subsection{End-to-end learning}
	With the growing popularity of deep learning, the traditional learning pipeline (i.e. pre-processing, feature extraction, modelling, inferencing) is replaced by a single \gls{DNN} which learns the features and performs the modelling in a data-driven manner. In this case, raw data or shallow features are directly fed into the learning process, and the \gls{DNN} model predicts the target (e.g. class labels, parameter estimates) at its output. This is referred as end-to-end learning. Due to the data-driven nature of end-to-end learning, very little domain expertise is used for learning, and brute-force feature validation is avoided, thereby significantly reducing labour costs. In addition, compared to the hand-crafted features that unavoidably leads to information loss during the extraction process, the end-to-end framework fuses feature learning and selection into one process so that dedicated features for the target problem can be learned, thus contributing to the modelling performance.
	
	In the context of \gls{SER}, Trigeorgis et al. first proposed an end-to-end \gls{DNN} that consists of \gls{CNN} layers for feature extraction from raw audio waveforms and \gls{LSTM} layers to model the temporal information in the feature space \cite{Trigeorgis2016}. In \cite{Sarma2018}, the authors replace the \gls{CNN} with a \gls{TDNN} to create a larger receptive field. A similar idea has been proposed in \cite{Tang2021} where a dilated \gls{CNN} having a receptive field as large as the input sequence is used for both feature extraction and temporal modelling. However, generalisability remains a problem in supervised end-to-end learning since annotated \gls{SER} datasets are general of small scale and recorded in specific environments, hence a model learned from these datasets may perform poorly when the test conditions are different from the training conditions, as well as in a cross-language setting.
	
	\subsection{Self-supervised learning and transfer learning}
	Self-supervised learning provides an end-to-end learning pathway for constructing speech features from large quantities of data without annotations. Self-supervised learning methods first transform time-domain raw audio waveforms to an embedding space (i.e. feature space), and then aim to predict randomly masked embeddings from past embeddings or adjacent embeddings \cite{VandenOordDeepMinda, Schneider2019, Baevski2020, Conneau2020}, or to predict the features generated though an augmented ``target'' network \cite{grill2020bootstrap}. These learning schemes force the \gls{DNN} to learn intrinsic contextual structure from the data rather than modelling noise or irrelevant factors, because on a large scale, noise will not contain useful information to predict neighbouring embeddings or augmented embeddings, thus will be suppressed.
	
	The speech features obtained from self-supervised learning can either be used directly in relevant tasks, or can be fine-tuned in a transfer learning framework. In the latter case, a few extra layers, which map the features into predictions, are added to the trained self-supervised feature extractor. Then, the predictions are evaluated with a supervised loss, and finally the entire set of \gls{DNN} model parameters is updated with a supervised training dataset. Examples in \gls{SER} are presented in \cite{Pepino2021, sharma2022multi, ma2024emotion2vec}.
	
	\section{Proposed transfer learning method with Deep-WCCN for cross-language \gls{SER}}
	\label{sec:xlser_method}
	The proposed method is based on the wav2vec 2.0 self-supervised learning model \cite{Baevski2020}. This model has been pre-trained and used in cross-language speech recognition and has been shown to deliver speech representations that are shared across languages \cite{Conneau2020}. We first give the details of the wav2vec 2.0 model in Section \ref{sec:xlser_wav2vec}, then we will present how to use the wav2vec 2.0 model in the front-stage of an end-to-end \gls{SER} system, together with a modified \gls{Deep-WCCN} layer inspired by similar work in \cite{Eghbal-zadeh2018} (in Section \ref{sec:xlser_deepwccn}) and finally how to fine-tune the model under a supervised scheme in Section \ref{sec:xlser_finetune}.
	
	\subsection{Self-supervised pre-training of wav2vec 2.0}
	\label{sec:xlser_wav2vec}
	The wav2vec 2.0 model learns contextual representations from speech waveforms. The model consists of a 
	convolutional feature extractor $f$, a transformer-based sequence model $e$ and a vector quantiser 
	module \cite{Conneau2020}. The feature extractor $f$ first transforms the raw speech waveform $\mathcal{X}$ 
	into a sequence of $d_z$-dimensional latent speech representations $\bm{Z} = [\bm{z}_1, \bm{z}_2, ..., \bm{z}_T]$ 
	for $T$ time-stamps. Then the latent speech representations on one hand are fed into the sequence 
	modeller to build contextual representations $\bm{c}_1, \bm{c}_2, ..., \bm{c}_T$, and on the other 
	hand are fed to the vector quantiser that contains $G$ codebooks, and for each codebook, there are 
	$V$ entries. The vector quantiser discretises the latent representations and linearly transforms them 
	into a sequence $\bm{q}_1, \bm{q}_2, ..., \bm{q}_T$. The feature extractor $f$, the sequence model $e$ and the 
	vector quantiser module are trainable and their parameters are optimised through back-propagation with two loss functions:
	\begin{enumerate}
		\item \emph{The contrastive loss:}
		For a $\bm{c}_t$ obtained from masked representations centered at time-stamp $t$, the sequence model needs to identify the true quantised representation $\bm{q}_t$ from $K+1$ candidate quantised representations $\bm{\tilde{q}}\in \bm{Q}_t$ which include $K$ distractors uniformly sampled from other masked time-stamps for the same sequence. The loss is defined as:
		\begin{equation}
			\mathcal{L}{\mathrm{_m}} = -\mathrm{log}\frac{\mathrm{exp}(\mathrm{sim}(\bm{c}_t, \bm{q}_t) /\kappa )}{\sum_{\tilde{\bm{q}}\in\bm{Q}_t} \mathrm{exp}(\mathrm{sim}(\bm{c}_t, \bm{\tilde{q}}_t) /\kappa )}
		\end{equation}
		where $\mathrm{sim}(\bm{a}, \bm{b})$ is the cosine similarity between $\bm{a}$ and $\bm{b}$, and $\kappa$ is the temperature parameter that controls the kurtosis of the distribution. 
		\item \emph{The diversity loss:} To increase the use of the codebooks, the diversity loss encourages the equal usage among the entries in every codebook by maximising the entropy {\color{myred} $H(\centerdot)$} of the averaged softmax distribution over codebook entries for each codebook $\bar{p}_g$ across a batch of utterances:
		\begin{equation}
			\mathcal{L}{\mathrm{_d}} = \frac{1}{GV}\sum_{g=1}^{G}-H(\bar{p}_g) =  \frac{1}{GV}\sum_{g=1}^{G}\sum_{v=1}^{V}\bar{p}_{g, v}\mathrm{log}(\bar{p}_{g, v})
		\end{equation}
	\end{enumerate}
	The total loss is a weighted summation of the two losses with a hyper-parameter $\alpha$,
	\begin{equation}
		\label{eq:xlser_originalloss}
		\mathcal{L}{\mathrm{_{ssl}}} = \mathcal{L}{\mathrm{_m}} + \alpha \mathcal{L}{\mathrm{_d}}
	\end{equation}
	
	To accommodate for cross-language speech variations, we use a pre-trained wav2vec 2.0 model that has been trained on 56,000 hours of speech in 53 different languages. This model is denoted as XLSR-53 and the details can be found in \cite{Conneau2020}.
	
	\subsection{\gls{Deep-WCCN}}
	\label{sec:xlser_deepwccn}
	The \gls{WCCN} was introduced in \cite{Hatch2006}, and used in speaker recognition and verification applications \cite{Hatch2006,Dehak2011}. The \gls{WCCN} approach aims to learn a feature map that minimises upper bounds on the false positive and false negative rates in a linear classifier. First, let $\bm{W}=[\bm{w}_1, \bm{w}_2, \dots, \bm{w}_N]$ denote a set of $N$ $d$-dimensional feature vectors belonging to $C$ different classes $c \in \{1, \dots, C\}$. In \cite{Dehak2011}, $\bm{W}$ is a set of i-vectors and $C$ is the number of different speakers. Differently in this paper, we define $\bm{W}$ to represent a set of intermediate features of the training set extracted from the front-end \gls{DNN} feature extractor, and $C$ is the total number of emotion classes.
	
	Next, we define the expected within-class covariance matrix as:
	\begin{equation}
		\bm{S}_w = \frac{1}{C}\sum_{c=1}^{C}\frac{1}{N_c}\sum_{i=1}^{N_c}(\bm{w}_i^c-\bm{\bar{w}^c})(\bm{w}_i^c-\bm{\bar{w}^c})^T
	\end{equation}
	where $\bm{\bar{w}^c}=\frac{1}{N_c}\sum_{i=1}^{N_c}\bm{w}_i^c$ is the mean feature vector of class $c$, and $\bm{w}_i^c$ are the examples belonging to this class. $N_c$ is the total number of examples belonging to class $c$ in the training set.
	
	Then the optimal feature map is given as:
	\begin{equation}
		\label{eq:xlser_wccnmap}
		\Phi(\bm{w}) = \bm{A}^T\bm{w}
	\end{equation}
	where $\bm{A}$ is the Cholesky factors of $\bm{S}_w^{-1}=\bm{A}\bm{A}^T$, and $\bm{w}$ represents an arbitrary feature vector.
	
	The conventional \gls{WCCN} pools the necessary statistics from the feature vectors of the entire training set, which is not compatible with mini-batch training in \glspl{DNN}. In \cite{Eghbal-zadeh2018}, the authors propose to estimate $\bm{S}_w$ with mini-batches, and to maintain a moving average $\hat{\bm{A}}$ of the corresponding mini-batch projection matrix. We believe that this moving average operation proposed in \cite{Eghbal-zadeh2018}, which is equivalent to a weighted average of new and old batches, where the old batches are subject to exponentially decaying weights, is not the best design choice for random mini-batch \gls{SGD} training in \glspl{DNN}. Therefore, we first define the class covariance matrix $\hat{ \bm{S}}_{w, c}$, which is estimated on mini-batch, and also $\hat{\bm{S}}_{w}$, which is the average of $\hat{ \bm{S}}_{w, c}$ across the classes. Then we propose to maintain a cumulative average $\bar{ \bm{S}}_{w}$ for $\hat{\bm{S}}_{w}$,
	\begin{equation}
		\label{eq:xlser_streamingavg}
		\bar{ \bm{S}}_{w} =  \frac{N\mathrm{_{tot}}}{N\mathrm{_{tot}}+1}\bar{\bm{S}}_{w} + \frac{1}{N\mathrm{_{tot}}+1}\hat{\bm{S}}_{w}
	\end{equation}
	where $N\mathrm{_{tot}}$ is an accumulator that counts the total number of mini-batches during training. Next, we add a spectral smoothing term to the within-class covariance estimation similar to \cite{Hatch2006}:
	\begin{equation}
		\label{eq:xlser_wccnsmooth}
		\bm{S}_w' = (1-\beta)\bar{\bm{S}}_w + \beta I
	\end{equation}
	where the hyper-parameter $\beta\in $[0, 1] controls the smoothness of the estimated within-class covariance matrix. Finally, $\bm{S}_w'$ is used to calculate $\bm{A}$ as mentioned before. 
	
	This mini-batch based \gls{WCCN} is considered as a special \gls{DNN} layer that has no trainable parameter and only updates $\bm{A}$ during training, denoted as \gls{Deep-WCCN}. The last update of $\bm{A}$ during training is stored and used in testing. The output of \gls{Deep-WCCN} is the result of the linear transformation in (\ref{eq:xlser_wccnmap}), which can be fed into the subsequent layers of the \gls{DNN} model. We implement the \gls{Deep-WCCN} using Pytorch, which is an automatic differentiation toolbox for deep learning. It should be noted that in order to avoid Pytorch to automatically generate gradients of $\bm{w}$ when calculating $\bm{A}$, a ``detached'' (from the computational graph) copy of $\bm{w}$ is used in the calculation during training.
	
	\subsection{Fine-tuning the wav2vec 2.0 model for \gls{SER}}
	\label{sec:xlser_finetune}
	
	\begin{figure}[t!]
		\centering
		\includegraphics[width=0.55\linewidth]{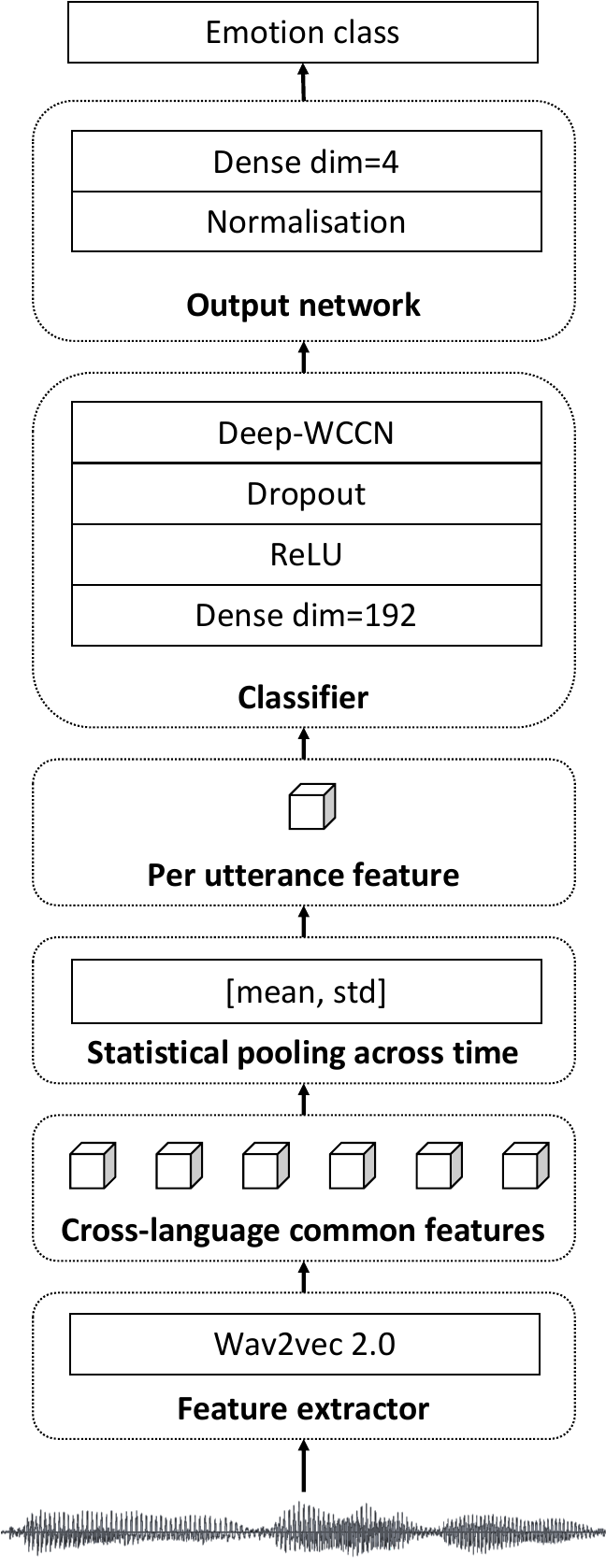}
		\caption{The network structure of the proposed end-to-end transfer learning model for cross-language/cross-corpus \gls{SER}.}
		\label{fig:xlser_network}
	\end{figure}
	
	One problem of the pre-trained XLSR-53 model is that the training sets mostly consist of neutral speech, and as a consequence the pre-trained model may capture inadequate emotion structures embedded in the speech data. Therefore, we propose to fine-tune the XLSR-53 model with emotional speech data. We combine the XLSR-53 pre-trained model and the \gls{Deep-WCCN} with a few extra layers into a unified model that maps time-domain raw waveforms to emotion class predictions, and this unified model is fine-tuned in a supervised manner. The model overview is shown in Figure\,\ref{fig:xlser_network}. After converting the speech raw waveform to the sequence of latent speech representations $\bm{Z}$, we calculate an utterance-level feature by pooling and concatenating the statistics of $\bm{Z}$ along the time dimension as done in \cite{Li2021},
	\begin{equation}
		\bm{u} = 
		\begin{bmatrix}
			\mathrm{mean}(\bm{Z}) \\
			\mathrm{std}(\bm{Z})
		\end{bmatrix}
	\end{equation}
	
	The resulting feature vector $\bm{u}$ is a $2d_z$-dimensional vector that proceeds to a fully-connected layer which reduces its dimensionality to a pre-defined hidden dimension $d\mathrm{_h}=\frac{1}{4} d_z$ and which is then followed by a \gls{ReLU} non-linear activation and a dropout layer. The output from the dropout layer is the intermediate feature vector that will be fed into the \gls{Deep-WCCN} to reduce the within-class variances. After that, the \gls{Deep-WCCN} output feature vectors are normalised to have unit norm, and linearly transformed into predictions for emotion classification.
	
	The fine-tuning loss is a weighted summation of the wav2vec 2.0 loss $\mathcal{L}\mathrm{_{ssl}}$ and the emotion classification log-softmax cross-entropy loss. Specifically, given one prediction $\bm{p}=[p_1, p_2, \dots, p_C]^T$ and its corresponding one-hot emotion class label $\bm{y}=[y_1, y_2, \dots, y_C]^T$ (where only the true class label is 1, and other labels are 0) the total loss is:
	\begin{equation}
		\label{eq:xlser_tot_loss}
		\mathcal{L}{\mathrm{_{tot}}} = -\sum_{c=1}^{C} y_c  \mathrm{log} \frac{\mathrm{exp} (p_c)}{\mathrm{exp} (\sum_{i=1}^{C} p_c)} + \gamma \mathcal{L}{\mathrm{_{ssl}}}
	\end{equation}
	where the hyper-parameter $\gamma$ controls the weight of $ \mathcal{L}{\mathrm{_{ssl}}}$.
	
	\section{Experiment set-ups}
	\label{sec:xlser_exp}
	
	\subsection{Datasets}
	To carry out simulations for cross-language \gls{SER}, we choose three publicly available datasets including English, German, and Chinese emotive speech recordings. These speech recordings are performed by professional actors have a duration of a few seconds each. The datasets are summarised in Table\,\ref{table:xlser_ds}, and described in detail below.
	
	\subsubsection{\gls{Emo-DB}}
	The \gls{Emo-DB} \cite{Burkhardt2005} dataset is a German emotive speech dataset that has been widely used in \gls{SER} research. This dataset consists of 535 utterances including 7 basic emotion catalogues (anger, boredom, disgust, anxiety/fear, happiness, sadness, and neutral). The utterances are performed by 10 professional actors, with 10 pre-defined sentences. Each sentence is performed with all different emotions, and the sentence content should not deliver sentimental information.
	
	\subsubsection{\gls{RAVDESS}}
	The \gls{RAVDESS} \cite{Livingstone2018} dataset contains recordings with 24 professional actors (12 female, 12 male), vocalising two lexically-matched statements (i.e. ``Kids are talking by the door.'', ``Dogs are sitting by the door.''). The speech recordings include calm, happy, sad, angry, fearful, surprise, and disgust expressions. Each expression is produced at two levels of emotional intensity (normal, strong), with an additional neutral expression, resulting in a total of 8 emotion catalogues.
	
	\subsubsection{\gls{ESD}}
	The \gls{ESD} \cite{Zhou2021} dataset is a recent multilingual and multi-speaker emotional speech dataset designed for various speech synthesis and voice conversion tasks. The dataset consists of 350 parallel utterances spoken by 10 native English and 10 native Mandarin speakers. In this work, we only use the Mandarin utterances, which involve 5 male speakers and 5 female speakers, and are performed in 5 emotion catalogues (happy, sad, neutral, angry, and surprise).
	
	\begin{table*}
		\begin{center}
			\begin{tabular}{P{0.14\linewidth} P{0.14\linewidth} P{0.14\linewidth} P{0.14\linewidth} P{0.14\linewidth} P{0.14\linewidth}}
				\toprule[1pt]
				& \textbf{\gls{Emo-DB}} & \textbf{\gls{RAVDESS}} & \textbf{\gls{ESD}} train & \textbf{\gls{ESD}} valid & \textbf{\gls{ESD}} test \\
				\hline
				Language & German & English & Chinese & Chinese & Chinese\\
				
				\# female speakers & 5 & 12 & 5 & 5 & 5\\
				
				\# male speakers & 5 & 12 & 5 & 5 & 5\\
				
				Reference & \cite{Burkhardt2005} & \cite{Livingstone2018} & \cite{Zhou2021} & \cite{Zhou2021} & \cite{Zhou2021} \\
				
				\hline
				\multicolumn{6}{c}{Utterance duration (s)} \\
				\hline
				
				Average & 2.54 & 1.72 & 2.47 & 2.41 & 2.49 \\
				
				Median & 2.3 & 1.66 & 2.3 & 2.3 & 2.37 \\
				
				Maximum & 8.88 & 3.41 & 7.07 & 5.6 & 5.66 \\
				
				Minimum & 0.83 & 0.86 & 0.42 & 0.9 & 0.9 \\
				
				Total & 860 & 1160 & 29634 & 1932 & 2990 \\
				
				\hline
				\multicolumn{6}{c}{Number of utterances per selected class} \\
				\hline
				
				Angry & 127 & 192 & 3000 & 200 & 300 \\
				
				Sad & 62 & 192 & 3000 & 200 & 300 \\
				
				Neutral & 79 & 96 & 3000 & 200 & 300 \\
				
				Happy & 71 & 192 & 3000 & 200 & 300 \\
				
				Total & 339 & 672 & 12000 & 800 & 1200 \\
				
				\bottomrule[1pt]
			\end{tabular}
			\caption{Dataset information.}
			\label{table:xlser_ds}
		\end{center}
	\end{table*}
	
	\subsubsection{Dataset preprocessing and partitioning}
	First, we only select 4 overlapping emotion catalogues (angry, happy, neutral, and sad) from the aforementioned datasets, and all recordings are re-sampled to 16\,kHz. Then, we zero-pad or randomly crop the time-domain raw waveforms to have a 2\,s duration. Next, we apply mean and variance normalisation across each waveform to match the requirements of the XLSR-53 pre-trained model.
	
	Finally, we partition each dataset into training, validation, and testing subsets. Each dataset is divided into 5 groups containing different speakers, which results in 2 speakers per group for the \gls{Emo-DB} dataset and the \gls{ESD} dataset, and 5 speakers for the first four groups of the \gls{RAVDESS} dataset and 4 speakers for the last group. To ensure there is no speaker overlap in the subsets, we use three groups for training, one group for validation, and one group for testing. This scheme results in 5 different partitionings of the datasets, and allows us to apply 5-fold cross-validation. The original speaker IDs used in validation and testing are listed in Table\,\ref{table:xlser_partitioning}, and the remaining speakers are used for training. Note that, the \gls{ESD} dataset has originally already partitioned into training, validation and testing sets for each speaker. Hence when creating \gls{ESD} subsets, we firstly merge the original training, validation and testing sets per speaker, then partition the new subsets again by speaker IDs.
	
	\begin{table*}[h!]
		\begin{center}
			\begin{tabular}{P{0.1\linewidth} P{0.14\linewidth} P{0.14\linewidth} P{0.14\linewidth} P{0.14\linewidth} P{0.14\linewidth}}
				\toprule[1pt]
				\textbf{Division} & \textbf{Fold 1} & \textbf{Fold 2} & \textbf{Fold 3} & \textbf{Fold 4} & \textbf{Fold 5} \\
				\hline
				\multicolumn{6}{c}{\gls{Emo-DB}} \\
				\hline
				
				
				Valid & 13, 14 & 11, 12 & 09, 10 & 03, 08 & 15, 16\\
				
				Test & 15, 16 & 13, 14 & 11, 12 & 09, 10 & 03, 08\\
				
				\hline
				\multicolumn{6}{c}{\gls{RAVDESS}} \\
				\hline
				
				Valid & 16, 17, 18, 19, 20 & 11, 12, 13, 14, 15 & 06, 07, 08, 09, 10 & 01, 02, 03, 04, 05 & 21, 22, 23, 24 \\
				
				Test & 21, 22, 23, 24 & 16, 17, 18, 19, 20 & 11, 12, 13, 14, 15 & 06, 07, 08, 09, 10 & 01, 02, 03, 04, 05 \\
				
				\hline
				\multicolumn{6}{c}{\gls{ESD}} \\
				\hline
				
				
				Valid & 0007, 0008 & 0005, 0006 & 0003, 0004 & 0001, 0002 & 0009, 0010\\
				
				Test & 0009, 0010 & 0007, 0008 & 0005, 0006 & 0003, 0004 & 0001, 0002\\
				
				\bottomrule[1pt]
			\end{tabular}
			\caption{Dataset partitioning for 5-fold cross-validation with reference to original speaker ID.}
			\label{table:xlser_partitioning}
		\end{center}
	\end{table*}
	
	\subsection{Baseline features}
	For comparison with the proposed model, we choose the baseline systems using the most widely used feature sets designed for \gls{SER}. The \gls{eGeMAPS} \cite{Eyben} and the Emobase \cite{Eyben2009} feature sets produce features per utterance, computed from a set of low-level-descriptors to which various statistical functions are applied. Both the \gls{eGeMAPS} and the Emobase feature sets contain spectral features (e.g., pitch-related, formant-related features), energy/amplitude features (e.g., loudness), and filter-bank features (e.g., \gls{MFCC}). In Emobase, statistical functions including min/max, arithmetic mean, standard deviation, skewness, and kurtosis, are applied to all the low-level descriptors and their delta coefficients. In contrast, in \gls{eGeMAPS}, only selected statistical functions are applied to some of the low-level descriptors, resulting a feature set containing a few extra temporal features such as the rate of loudness peaks features, and the mean length of voiced regions and unvoiced regions. The low-level descriptor selection criteria for \gls{eGeMAPS} are based on the theoretical significance of the feature, the feature usage frequency in literature,  and the potential of an acoustic parameter to indicate physiological changes in voice production during affective processes \cite{Eyben}. In term of the feature vector dimensionality, the Emobase feature vector has length 988, and the \gls{eGeMAPS} feature vector has length 88.
	
	\subsection{Training on multiple languages}
	Similarly to \cite{Latif2018,Gerczuk2021}, we consider all combinations of merging two out of three language datasets for training and validation in order to meet the data requirements of the data-driven model and to allow the model to capture the variations across languages and corpora thus not to over-fit to one dataset.
	
	In combination of each two datasets, we repeat the smallest dataset a few times so that for each training set, there is a similar amount of utterances from each language. This results in the following three training/validation sets:
	\begin{enumerate}
		\item \textbf{DECH}: training and validation using German and Chinese recordings, repeating the German training set 32 to 39 times depending on which fold is considered.
		\item \textbf{DEEN}: training and validation using German and English recordings, repeating the German training set 2 to 3 times depending on which fold is considered.
		\item \textbf{ENCH}: training and validation using Chinese and English recordings, repeating the English training set 18 to 19 times depending on which fold is considered.
	\end{enumerate}
	
	\subsection{Within-language and Cross-language settings}
	\subsubsection{Within-language \gls{SER}}
	In the within-language setting, we aim to evaluate the effectiveness of the baseline features and the wav2vec 2.0 features in \gls{SER}. Therefore, for each training scheme, we test the performance of the methods on the testing set in one of the training languages, resulting in a situation where only the speaker identity is different among the training and testing sets.
	
	\subsubsection{Cross-language \gls{SER}}
	To evaluate the methods in a cross-language setting, we employ a leave-one-language-out scheme, that is for each training scheme, we test the methods on the testing set of the third language which means the language, recording condition, and the speaker identity in the testing set is unseen during training. 
	
	We use ``-{\textbf \footnotesize \textgreater}'' followed by the language abbreviations (EN, CH, DE) to indicate the dataset/language used in the testing.
	
	\subsection{Experimental settings}
	The Adagrad optimiser \cite{Duchi2010} is used to train the models, with a fixed learning rate of $3\times10^{-4}$ and weight decay. The batch size is 14 due to memory constraints. The weighting parameter $\alpha$ in (\ref{eq:xlser_originalloss}) is equal to 0.1 as in \cite{Conneau2020}. The weight decay parameter, the dropout rate, and the hyper-parameters $\beta$ and $\gamma$ in (\ref{eq:xlser_wccnsmooth}) and (\ref{eq:xlser_tot_loss}) are optimised using Hyperopt \cite{Bergstra2013}. The optimal results are shown in Table\,\ref{table:xlser_hyperparam} and correspond to the values that are used in the experiments. 
	
	We evaluate the model using two metrics, the \gls{UA} and the \gls{WA}. Specifically, \gls{UA} is the average accuracy for each emotion class, which marginalises out the effect of class imbalance, and \gls{WA} is the overall accuracy of the entire testing data, which indicates the overall model performance across all classes. Model selection uses early-stopping on the validation performance, and the reported testing results are averaged across cross-validation folds.
	
	\subsection{Baseline classifier}
	The baseline features are firstly tested with a neural network model that has the same structure as the classifier and the output network in the proposed model shown in Figure\,\ref{fig:xlser_network}. However, this model essentially only contains one non-linear transformation and the \gls{Deep-WCCN} operation, hence it is incapable to learn a good mapping from the baseline features to the emotion classes. In particular, when conducting a hyper-parameter search for the \gls{Deep-WCCN} layer, the dropout rate, and the weight decay rate using Hyperopt for the neural network model, and then training it with the baseline features, the model \gls{SER} performance is close to chance level (e.g., 25\% \gls{WA} for 4 emotion classes). Therefore, we instead use a \gls{RF} classifier, which is essentially an ensemble classifier that has good generalisation capability. This \gls{RF} classifier has 100 trees with maximum tree-depth of 5. 
	
	\begin{table}[h!]
			\begin{tabular}{P{0.3\linewidth} P{0.17\linewidth} P{0.17\linewidth} P{0.17\linewidth}}
				\toprule[1pt]
				\textbf{Hyper-paramter} & \textbf{DECH} & \textbf{DEEN} & \textbf{ENCH} \\
				\hline
				Learning rate & $3\times 10^{-4}$ & $3\times 10^{-4}$ & $3\times 10^{-4}$ \\
				
				Weight decay & $4\times 10^{-4}$ & $5\times 10^{-4}$ & $5\times 10^{-4}$ \\
				
				Dropout & 0.45 & 0.05 & 0.3 \\
				
				$\beta$ & 0.2 & 0.5 & 0.4 \\
				
				$\gamma$ & $8\times 10^{-4}$& $1.2\times 10^{-3}$& $1.7\times 10^{-3}$ \\
				
				Batch size & 14 & 14 & 14 \\
				
				\bottomrule[1pt]
			\end{tabular}
		\caption{Hyper-parameter configurations for the proposed method.}
		\label{table:xlser_hyperparam}
	\end{table}
	
	\begin{figure*}[h!]
		\centering
		\includegraphics[width=\linewidth]{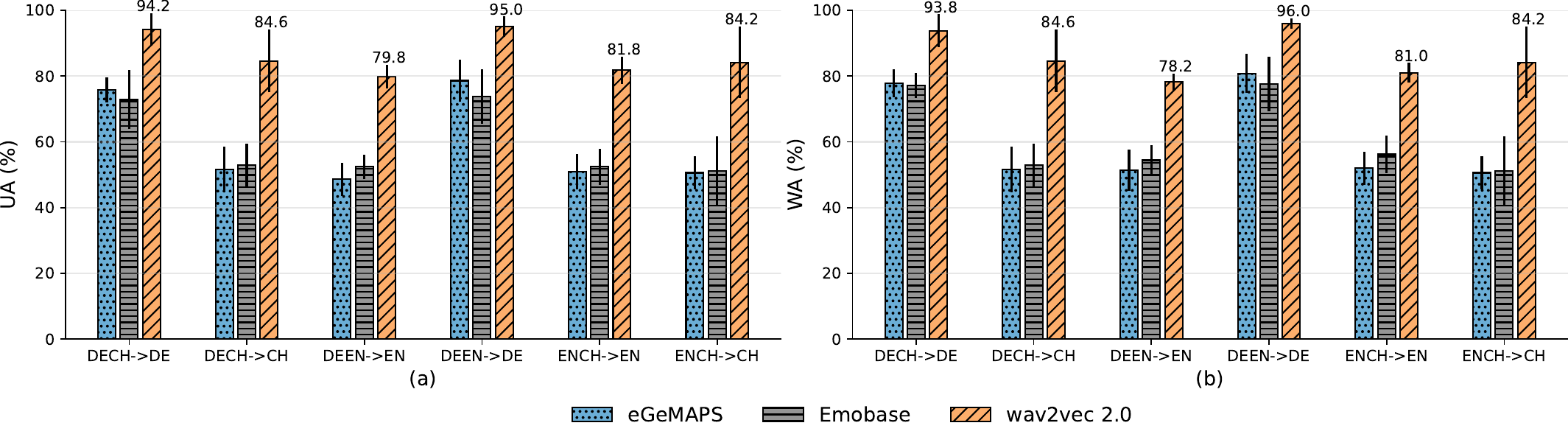}
		\caption{Within-language emotion classification accuracy measured by unweighted (\gls{UA}) and weighted accuracy. The training and validation sets are in the languages indicated on the left hand side of -\textgreater, and the testing set is in the language indicated on the right hand side of -\textgreater. The testing set language is included in the training, but the testing speaker ID is unseen.}
		\label{fig:xlser_exp1}
	\end{figure*}
	
	\section{Results and discussion}
	\label{sec:xlser_results}
	\subsection{Within-language performance}
	\label{sec:xlser_withinlangresults}
	The results for within-language evaluation are illustrated in Figure\,\ref{fig:xlser_exp1}. These results show that the baseline \gls{eGeMAPS} feature and the Emobase feature with an \gls{RF} classifier perform similarly in terms of \gls{UA} and \gls{WA}, and for some testing cases (e.g. testing on the German ``-{\footnotesize \textgreater}DE'' dataset) the \gls{eGeMAPS} performs better than the Emobase, but it is the other way around for the other testing cases (e.g. testing on the Chinese ``-{\footnotesize \textgreater}CH'', and on the English ``-{\footnotesize \textgreater}EN'' datasets). The largest \gls{UA} and \gls{WA} differences between these two methods are 4.8\% and 4.2\% that are obtained in the ``DEEN-{\footnotesize \textgreater}DE'' and ``ENCH-{\footnotesize \textgreater}EN'' scenarios, respectively. 
	
	From the same figure, comparing the proposed wav2vec 2.0 method with the baseline methods, the wav2vec 2.0 shows significant improvements in both \gls{UA} and \gls{WA} for within-language \gls{SER}. Specifically, the wav2vec 2.0 improves \gls{UA} with about 20.9\% to 66.4\% compared to the \gls{eGeMAPS} method, and with about 28.7\% to 64.5\% compared to the Emobase method, and it improves \gls{WA} with about 18.8\% to 66.4\% and 21.5\% to 64.5\% compared to the \gls{eGeMAPS} and the Emobase, respectively. The increase in classification accuracy may indicate that the wav2vec 2.0 pre-trained feature extractor combined with the proposed \gls{Deep-WCCN} can extract contextual features that contribute to mixed-language \gls{SER} capability and are robust to varying speaker identities. 
	
	\subsection{Cross-language/cross-corpus performance}
	\begin{figure*}[h!]
		\centering
		\includegraphics[width=\linewidth]{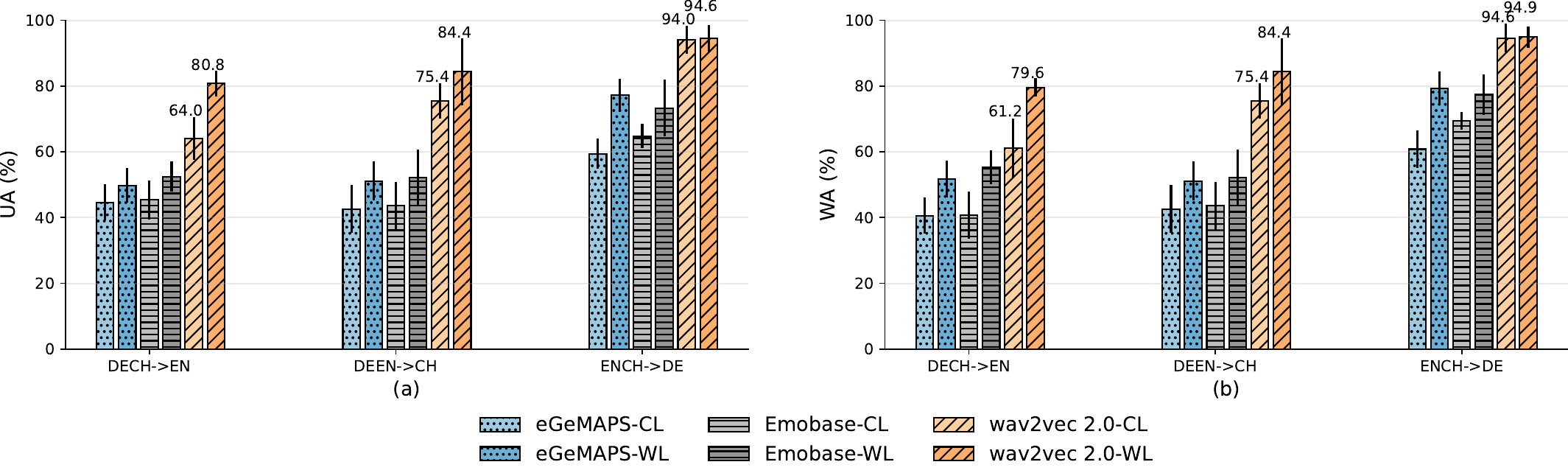}
		\caption{Cross-language/cross-corpus emotion classification accuracy measured by unweighted (\gls{UA}) and weighted accuracy. The testing set language is not included in training, which means the testing speaker ID, the testing language, and the testing channel is unseen during training. Results are compared also with the within-language performance averaged across each testing language.}
		\label{fig:xlser_exp2}
	\end{figure*}
	
	For the cross-language/cross-corpus experiments, the emotion classification accuracy is plotted in Figure\,\ref{fig:xlser_exp2}. For comparison, the within-language results (with suffix ``-WL'') from Section\,\ref{sec:xlser_withinlangresults} are averaged across the same testing language for every method and plotted along with the cross-language performance (with suffix ``-CL''). 
	
	Firstly, the wav2vec 2.0-CL performs significantly better than \gls{eGeMAPS}-CL and Emobase-CL in all three testing languages, and it is followed by Emobase-CL which performs slightly better than \gls{eGeMAPS}-CL in almost all the testing cases (except on English testing where \gls{eGeMAPS}-CL and Emobase-CL have similar \gls{WA} performance). The best \gls{UA} and \gls{WA} with wav2vec 2.0-CL, tested in English, Chinese, and German, are 64\%, 75.4\%, 94\% and 61.2\%, 75.4\%, 94.6\%, respectively. This is about 43\% to 77\% and 50.7\% to 77\% performance gain on \gls{UA} and \gls{WA} compared to the \gls{eGeMAPS}-CL, and about 41\% to 72.9\% and 36.3\% to 72.9\% performance gain on \gls{UA} and \gls{WA} compared to the Emobase-CL.
	
	Secondly, although all three methods show performance degradation when moving from within-language to cross-language/cross-corpus \gls{SER}, the wav2vec 2.0 still maintains high performance. It is notable that in the ``ENCH-{\footnotesize \textgreater}DE'' experiment, wav2vec 2.0-CL shows a very subtle degradation (i.e. about 0.6\% and 0.3\% degradation in \gls{UA} and \gls{WA}, respectively, compared to the wav2vec 2.0-WL), whereas the \gls{eGeMAPS}-CL shows a 17.8\% and 18.5\% decrease in \gls{UA} and \gls{WA}, respectively, and the Emobase-CL shows a 8.5\% and 8\% decrease in \gls{UA} and \gls{WA}, respectively.
	
	In summary, the results in cross-language/cross-corpus \gls{SER} may indicate that the proposed wav2vec 2.0 with \gls{Deep-WCCN} model can largely alleviate the performance degradation due to language mis-match, speaker identity mis-match, and channel mis-match that commonly occur in cross-language \gls{SER}. As the proposed model shows equally satisfactory results in both \gls{UA} and \gls{WA}, we may also conclude that the wav2vec 2.0 with \gls{Deep-WCCN} model is not over-fitting to one of the testing emotion classes.
	
	\subsection{Evaluation of the \gls{Deep-WCCN}}
	\begin{figure*}[h]
		\centering
		\includegraphics[width=\linewidth]{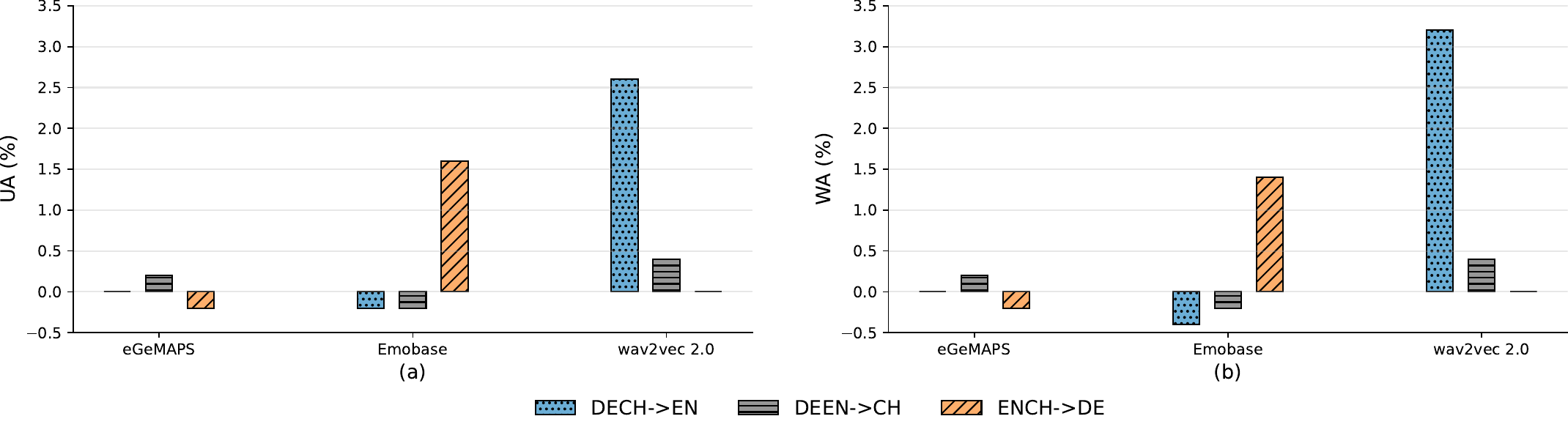}
		\caption{The performance gain in terms of \gls{UA} and \gls{WA} after applying \gls{Deep-WCCN} for the proposed method (which is based on wav2vec 2.0), and the performance gain after applying conventional \gls{WCCN} for \gls{eGeMAPS} and Emobase methods.}
		\label{fig:xlser_exp3}
	\end{figure*}
	
	\begin{figure*}[h!]
		\centering
		\includegraphics[width=\linewidth]{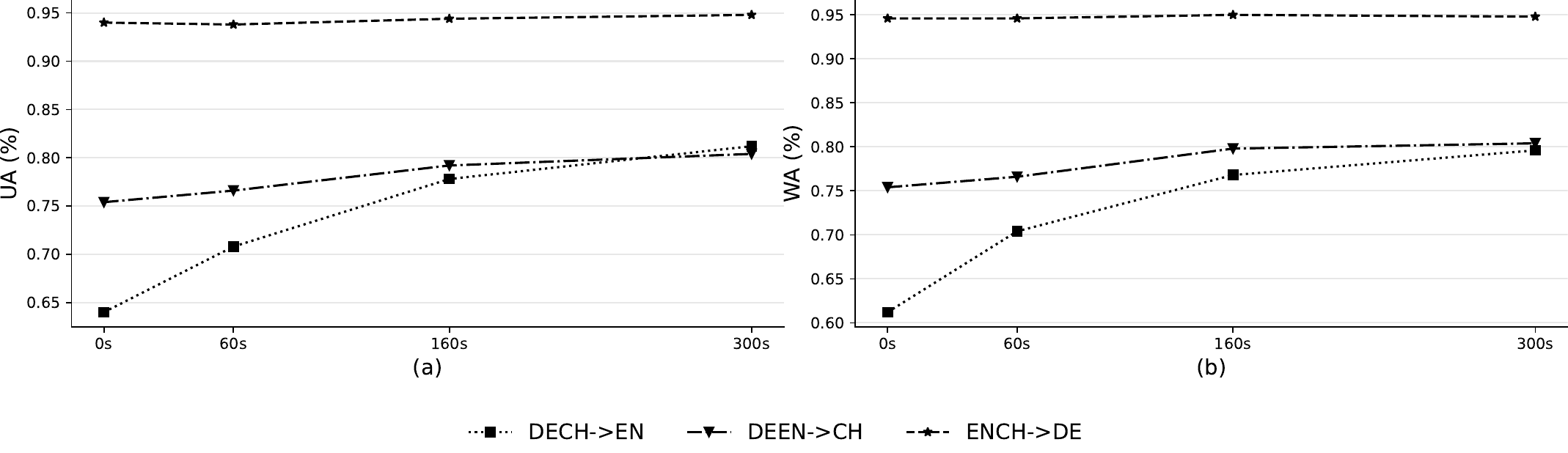}
		\caption{The \gls{SER} performance when adding additional target language data (i.e. 60\,s, 160\,s and 300\,s of target language recording) into training. Testing speaker IDs are still unseen during training.}
		\label{fig:xlser_exp4}
	\end{figure*}
	
	\begin{figure*}[htp]
		\centering
		\includegraphics[width=1.1\linewidth]{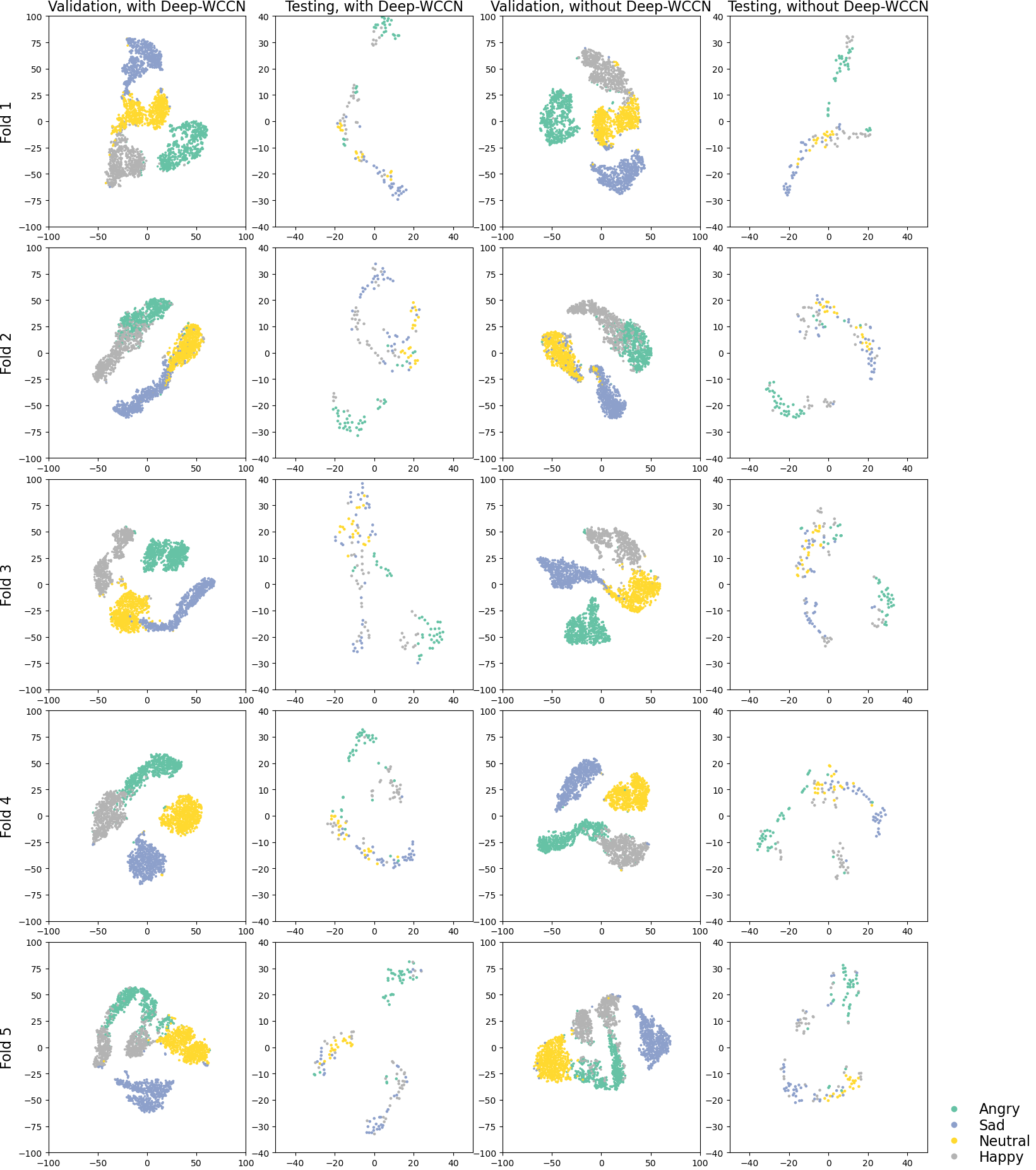}
		\caption{The \gls{t-SNE} visualisation of validation set and testing set embeddings when applying \gls{Deep-WCCN} and not applying Deep-\gls{WCCN}.}
		\label{fig:xlser_emb}
	\end{figure*}
	
	In order to verify the effectiveness of \gls{Deep-WCCN}, we compare the change in cross-language \gls{SER} performance of wav2vec 2.0 with and without \gls{Deep-WCCN}. The results are compared with other baseline methods. For \gls{eGeMAPS} and Emobase, we calculate the change in their cross-language \gls{SER} performance with and without \gls{WCCN}. The increase in cross-language \gls{SER} performances for all methods using either \gls{Deep-WCCN}  or \gls{WCCN} are shown in Figure\,\ref{fig:xlser_exp3}.
	
	First, it can be clearly observed that using \gls{Deep-WCCN} can improve the performance of wav2vec 2.0 in cross-language \gls{SER}. The highest increase is observed in DECH-{\footnotesize \textgreater}EN, where \gls{UA} and \gls{WA} increase by 2.6\% and 3.2\% respectively by applying \gls{Deep-WCCN}, which is followed by increments in DEEN-{\footnotesize \textgreater}CH, where both \gls{UA} and \gls{WA} increase by 0.4\% after applying \gls{Deep-WCCN}. No performance gain is observed in ENCH-{\footnotesize \textgreater}DE when applying \gls{Deep-WCCN} to the wav2vec 2.0 model, which might be due to the fact that the results on the German testing sets are already very good and there is limited room for improvement.
	
	Secondly, \gls{WCCN} did not significantly improve cross-language \gls{SER} performance when using the \gls{eGeMAPS} and the Emobase features. Specifically, by applying \gls{WCCN}, the experiments on all three cross-language cases show only small performance gains, no gains, or even performance drops when applying \gls{WCCN} to \gls{eGeMAPS} and to Emobase. Only in the case of ENCH-{\footnotesize \textgreater}DE, the Emobase method has increases of 1.6\% and 1.4\% in \gls{UA} and \gls{WA}, respectively, when applying \gls{WCCN}. This is due to the fact that \gls{WCCN} is designed for linear classifiers and is not significantly helpful for ensemble models like the \gls{RF}. Conversely, the output layer of our proposed model can be seen as a linear classifier, which satisfies the design conditions of \gls{WCCN}.
	
	We also visualise the embeddings (which are the inputs to the output network) with and without \gls{Deep-WCCN}. The embeddings are originally 192-dimensional vectors, but for visualisation purposes, we compute \glspl{t-SNE} on the validation set and testing set embeddings from each cross-validation fold. The visualisations for the DECH setting are plotted in Figure\,\ref{fig:xlser_emb}.
	
	First, distinct clusters corresponding to different emotion classes emerge, demonstrating the model's ability to learn meaningful representations that separate the emotion categories. This explains the proposed method's effectiveness in \gls{SER}. Second, incorporating \gls{Deep-WCCN} notably improves the purity of the clusters in the validation sets. This improvement in cluster purity indicates a better discrimination of emotion classes when using \gls{Deep-WCCN}, suggesting that the embeddings retain a better separability than without \gls{Deep-WCCN}, even when applied to unseen data. Third, the proposed method generalises well to testing data that comes from a different language. And by applying \gls{Deep-WCCN}, the embeddings exhibit enhanced discrimination properties, which help maintain robust performance in cross-language \gls{SER}.
	
	\subsection{Influence of training set size of target language}
	\begin{table*}[ht]
		\begin{center}
			\begin{tabular}{p{0.2\linewidth}p{0.17\linewidth}p{0.15\linewidth}p{0.11\linewidth}P{0.08\linewidth}p{0.06\linewidth}p{0.06\linewidth}}
				\toprule[1pt]
				\textbf{References} & \textbf{Method} & \textbf{Training language} & \textbf{Testing dataset} & \textbf{\# classes} & \textbf{WA} & \textbf{UA} \\
				\hline
				\multicolumn{7}{c}{Within-language \gls{SER}} \\
				\hline
				
				\multirow{3}{\linewidth}{Rehman et. al. \cite{Rehman2020}} & \multirow{3}{\linewidth}{Hybrid \gls{NN}} & EN+DE & \gls{RAVDESS} & 4 & - & 56.2\% \\
				&  & EN+EN & \gls{RAVDESS} & 4 & - & 62.5\% \\
				&  & EN+DE & \gls{Emo-DB} & 4 & - & 61.2\% \\
				\hline
				Latif et. al. \cite{Latif2019} & Domain adaptation & DE & \gls{Emo-DB} & 5 & - & 81.3\% \\
				\hline
				\multirow{2}{\linewidth}{Parry et. al. \cite{Parry2019}} & \multirow{2}{\linewidth}{Datasets aggregation} & EN+IT+DE & \gls{Emo-DB} & 3 & - & 69.72\% \\
				& & EN+IT+DE & \gls{RAVDESS} & 3 & - & 65.67\% \\
				\hline
				Desplanques and Demuynck \cite{Desplanques2018} & \multirow{2}{\linewidth}{Emotion i-vector} & DE & \gls{Emo-DB} & 4 & 90.5\% & - \\
				\hline
				\multirow{2}{\linewidth}{Proposed} & \multirow{2}{\linewidth}{Transfer learning + \gls{Deep-WCCN}} & EN+CH  EN+DE & \gls{RAVDESS} & 4 & \textbf{79.6\%} & \textbf{80.8\%} \\
				&  & EN+DE  CH+DE & \gls{Emo-DB} & 4 & \textbf{94.9\%} & \textbf{94.6\%}\\
				
				\hline
				\multicolumn{7}{c}{Cross-language \gls{SER}} \\
				\hline
				
				Rehman et. al. \cite{Rehman2020} & Hybrid \gls{NN} & EN+EN & \gls{Emo-DB} & 4 & - & 40.2\% \\
				\hline
				\multirow{2}{\linewidth}{Zhang et. al. \cite{Zhang2020}} & \multirow{2}{\linewidth}{Subspace learning} & EN & \gls{Emo-DB} & 5 & 47.35\% & - \\
				&  & TU & \gls{Emo-DB} & 5 & 49.26\% & - \\
				\hline
				Latif et. al. \cite{Latif2019} & Domain adaptation & IT+UR+EN & \gls{Emo-DB} & 5 & - & 68\% \\
				\hline
				Parry et. al. \cite{Parry2019} & Datasets aggregation & EN & \gls{Emo-DB} & 3 & - & 41.99\% \\
				\hline
				Desplanques and Demuynck \cite{Desplanques2018} & \multirow{2}{\linewidth}{Emotion i-vector} & EN & \gls{Emo-DB} & 4 & 81.4\% & - \\
				\hline
				Ma et. al. \cite{ma2024emotion2vec} & Self-supervised learning & EN+CH+IT+ FR+BN+PS+ UR+GR+RU & \gls{Emo-DB} & 4 & 84.34\% & 84.85\% \\
				\hline
				\multirow{2}{\linewidth}{Proposed} & \multirow{2}{\linewidth}{Transfer learning + \gls{Deep-WCCN}} & EN+CH & \gls{Emo-DB} & 4 & \textbf{94.6\%} & \textbf{94\%} \\
				&  & DE+CH & \gls{RAVDESS} & 4 & \textbf{61.2\%} & \textbf{64\%} \\
				& & & & & & \\
				
				\bottomrule[1pt]
			\end{tabular}
			\caption{Comparison to recent work in cross-language/cross-corpus \gls{SER}. EN, DE, IT, CH, TU, FR, BN, PS, GR, RU and UR indicate English, German, Italian, Chinese, Turkish, French, Bengali, Persian, Greek, Russian and Urdu language, respectively.}
			\label{table:xlser_stoa}
		\end{center}
	\end{table*}
	
	As the proposed method is in the realm of transfer learning, it might exhibit the data efficiency property typically associated to transfer learning. That is, the transferred sub-network, which is trained on a large quantity of speech data, learns intrinsic speech structure and can easily generalise to similar tasks with a small amount of annotated target data. Therefore, in this experiment, we merge a small amount of target language data into the fine-tuning process, and evaluate the model performance when using 30, 80, and 150 two-second-target language inputs with their corresponding labels. The duration of extra target language data used in the training is hence equal to 60\,s, 160\,s, and 300\,s in total duration, respectively. The extra target data are repeated to achieve a size that is similar to the original training data size to avoid data imbalance for different languages. The results for DECH-\textgreater EN, DEEN-\textgreater CH and ENCH-\textgreater DE are plotted respectively in Figure \ref{fig:xlser_exp4}, together with their performance when no extra target language data is used (i.e. the case corresponding to 0\,s).
	
	First, the three testing cases all show an increasing trend when more target language data is used in training, and their performance increases rapidly when even less than 160\,s of target language data is used while it slows down when even more target language data is used. Second, the performance of DECH-\textgreater EN increases the most, followed by DEEN-\textgreater CH, and lastly by ENCH-\textgreater DE. For the case of using 160\,s of target language data, compared to no target language data being used in training,  DECH-\textgreater EN, DEEN-\textgreater CH and ENCH-\textgreater DE perform 13.8\%, 3.8\% and 0.4\% better in \gls{UA}, respectively, and 15.6\%, 4.4\% and 0.4\% better in \gls{WA}, respectively. The low performance increment in ENCH-\textgreater DE might be due to the limited room for improvement. Third, except for ENCH-\textgreater DE having limited improvement, DEEN-\textgreater CH also has overall less improvement in \gls{UA} compared to DECH-\textgreater EN, when 300\,s of target language data is used in training. This could be due to the difficulty of transferring Germanic languages to Sino-Tibetan languages, which probably can be alleviated by using more target language annotated data, however this hypothesis might need more investigation.
	
	\subsection{Comparison to existing works}
	We also compare our method with a few recent methods proposed for speaker-independent cross-language/cross-corpus \gls{SER}. The comparison is in Table\,\ref{table:xlser_stoa} where we only present the results reported on the same testing datasets (i.e. on the \gls{Emo-DB} and \gls{RAVDESS}) as our work. Since there is hardly any research using \gls{ESD} for cross-language \gls{SER}, we do not include the results for this dataset. 
	
	The comparison shows that our proposed method has significantly improved the \gls{SER} performance in both within-language and cross-language scenarios with both \gls{Emo-DB} and \gls{RAVDESS} datasets. This may indicate that transfer learning with pre-training on large speech data and \gls{Deep-WCCN} plays an important role in \gls{DNN} generalisation for \gls{SER}.
	
	\section{Conclusions and future work}
	\label{sec:xlser_conclusions}
	To alleviate the performance degradation in cross-language/cross-corpus \gls{SER} compared to within-language/within-corpus \gls{SER}, we proposed a transfer learning method that firstly uses the wav2vec 2.0 pre-trained model to transfer a time-domain audio waveform into a contextual embedding space that is shared across different languages, thereby reducing the language variabilities in the speech features. Then, by applying a \gls{Deep-WCCN} layer, which is adapted to cope with \gls{DNN} training, this \gls{Deep-WCCN} layer can further reduce within-class variance caused by other factors (e.g. speaker identity, channel variability). Experimental results first show that the proposed method largely increases both within-language and cross-language \gls{SER} performance compared to the \gls{eGeMAPS} and Emobase feature sets that have been designed for and widely used in \gls{SER}. Furthermore, an ablation study shows that \gls{Deep-WCCN} can reduce the within-class variances which further improves the performance for the proposed \gls{DNN} model. In contrast, the conventional \gls{WCCN} does not show improvements on the \gls{eGeMAPS} and Emobase feature sets with a \gls{RF} classifier. Next, we show that the proposed transfer learning method exhibits good data efficiency in merging target language data in the fine-tuning process. The model speaker-independent \gls{SER} performance increases for all testing target languages, and a performance gain in \gls{WA} up to 15.6\% and in \gls{UA} up to 13.8\% is achieved when only 160\,s of target language data is used in the training set for fine-tuning. Finally, a comparison with recent work in cross-language/cross-corpus \gls{SER} demonstrates that the proposed method can significantly improve speaker-independent within-language and cross-language \gls{SER} performance among multiple datasets. 
	
	Future work firstly include the evaluation of the wav2vec 2.0 model pre-trained on an even larger speech dataset and the assessment of its benefit to cross-language \gls{SER} performance. A good candidate dataset is proposed in \cite{Babu2021} and consist of half a million hours of publicly available speech audio in 128 languages. Second{\color{myred}ly}, we plan to evaluate the proposed method on more emotive speech datasets, and to fine-tune the model on emotive speech datasets from more languages to create a general data-driven model for \gls{SER}. {\color{myred} Thirdly, a fine-tuning incorporating rank decomposition matrices as proposed in \cite{hu2022lora} could be tested under the scope of this paper. The fine-tuning scheme in \cite{hu2022lora} is expecting to retain the knowledge captured by the trained models, therefore improving the efficiency when adapting the model to more emotive speech datasets.} Lastly, it may be relevant to reduce the model size through model distillation \cite{Hinton2015} to make it feasible for mobile and low-power devices.
	
	\backmatter

	\section*{Declarations}
	\subsection*{Availability of data and materials}
	The datasets generated and/or analysed during the current study are available in the \gls{Emo-DB} repository (\url{http://emodb.bilderbar.info/docu/#emodb}), the \gls{RAVDESS} repository (\url{https://www.kaggle.com/datasets/uwrfkaggler/ravdess-emotional-speech-audio}) and the \gls{ESD} repository (\url{https://github.com/HLTSingapore/Emotional-Speech-Data}). 
	
	\subsection*{Competing interests}
	The authors declare that they have no competing interests.
	
	\subsection*{Funding}
	This research work was carried out at the ESAT Laboratory of KU Leuven. The research leading to these results has received funding from the Chinese Scholarship Council (CSC) (grant no. 201707650021), the KU Leuven Internal Funds C2-16-00449 and VES/19/004, and the European Research Council under the European Union's Horizon 2020 research and innovation program / ERC Consolidator Grant: SONORA (no.\,773268). This paper reflects only the authors' views and the Union is not liable for any use that may be made of the contained information.
	
	\subsection*{Authors' contributions}
	DWT invented the concept of transfer learning using the wave2vec 2.0 model to eliminate the language variability in cross-language speech emotion recognition. Then DWT proposed a \gls{Deep-WCCN} layer to compensate for other variabilities (i.e. other than language variabilities). DWT, PK, LG, and TvW jointly developed the research methodology to turn this concept into a usable and effective algorithm. DWT and TvW jointly designed and interpreted the computer simulations. DWT implemented the computer simulations and managed to the dataset used. All authors contributed in writing the manuscript, and further read and approved the final manuscript.
	
	\subsection*{Acknowledgements}
	Not applicable.

	\bibliography{literature_all}

@inproceedings{hu2022lora,
	title={Lo{RA}: Low-Rank Adaptation of Large Language Models},
	author={Hu, Edward J and Shen, Yelong and Wallis, Phillip and Allen-Zhu, Zeyuan and Li, Yuanzhi and Wang, Shean and Wang, Lu and Chen, Weizhu},
	booktitle={Proc. of 10th Int. Conf. Learn. Represent. (ICLR 2022)},
	year={2022},
}

@article{Reynolds1995,
	author = {Reynolds, Douglas A. and Rose, Richard C.},
	journal = {IEEE Trans. Speech Audio Process.},
	number = {1},
	pages = {72--83},
	title = {{Robust Text-Independent Speaker Identification Using Gaussian Mixture Speaker Models}},
	volume = {3},
	year = {1995}
}

@inproceedings{Desplanques2018,
	author = {Desplanques, Brecht and Demuynck, Kris},
	booktitle = {Proc. Annu. Conf. Int. Speech Commun. Assoc. (INTERSPEECH 2018)},
	pages = {3648--3652},
	title = {{Cross-lingual speech emotion recognition through factor analysis}},
	month = {Sep.},
	year = {2018}
}

@article{Dehak2011,
	author = {Dehak, Najim and Kenny, Patrick J. and Dehak, R{\'{e}}da and Dumouchel, Pierre and Ouellet, Pierre},
	journal = {IEEE Trans. Audio, Speech Lang. Process.},
	number = {4},
	pages = {788--798},
	publisher = {IEEE},
	title = {{Front-end factor analysis for speaker verification}},
	volume = {19},
	year = {2011}
}

@inproceedings{Zhao2019a,
	author = {Zhao, Ziping and Bao, Zhongtian and Zhang, Zixing and Cummins, Nicholas and Wang, Haishuai and Schuller, Bj{\"{o}}rn},
	booktitle = {Proc. of Annu. Conf. Int. Speech Commun. Assoc. (INTERSPEECH 2019)},
	pages = {206--210},
	title = {{Attention-enhanced connectionist temporal classification for discrete speech emotion recognition}},
	month = {Sep.},
	year = {2019}
}

@inproceedings{Trigeorgis2016,
	author={G. {Trigeorgis} and F. {Ringeval} and R. {Brueckner} and E. {Marchi} and M. A. {Nicolaou} and B. {Schuller} and S. {Zafeiriou}},
	booktitle={Proc. of 2016 IEEE Int. Conf. Acoust., Speech, Signal Process. (ICASSP '16)},
	title={Adieu features? End-to-end speech emotion recognition using a deep convolutional recurrent network},
	year={2016},
	volume={},
	number={},
	pages={5200-5204},
}

@inproceedings{Sarma2018,
	author = {Sarma, Mousmita and Ghahremani, Pegah and Povey, Daniel and Goel, Nagendra Kumar and Sarma, Kandarpa Kumar and Dehak, Najim},
	booktitle = {Proc. of Annu. Conf. Int. Speech Commun. Assoc. (INTERSPEECH 2018)},
	pages = {3097--3101},
	title = {{Emotion identification from raw speech signals using DNNs}},
	month = {Sep.},
	year = {2018}
}

@inproceedings{Schuller2009,
	author = {Schuller, Bj{\"{o}}rn and Vlasenko, Bogdan and Eyben, Florian and Rigoll, Gerhard and Wendemuth, Andreas},
	booktitle = {Proc. of 2009 IEEE Work. Autom. Speech Recognit. Understanding (ASRU 2009)},
	pages = {552-557},
	title = {{Acoustic emotion recognition: A benchmark comparison of performances}},
	year = {2009}
}

@article{Eyben,
	author={F. {Eyben} and K. R. {Scherer} and B. W. {Schuller} and J. {Sundberg} and E. {André} and C. {Busso} and L. Y. {Devillers} and J. {Epps} and P. {Laukka} and S. S. {Narayanan} and K. P. {Truong}},
	journal={IEEE Trans. Affect. Comput.},
	title={The Geneva Minimalistic Acoustic Parameter Set (GeMAPS) for Voice Research and Affective Comput.},
	year={2016},
	volume={7},
	number={2},
	pages={190-202},
}

@inproceedings{Tzirakis2018,
	author={P. {Tzirakis} and J. {Zhang} and B. W. {Schuller}},
	booktitle={Proc. of 2018 IEEE Int. Conf. Acoust., Speech, Signal Process. (ICASSP '18)},
	title={End-to-End Speech Emotion Recognition Using Deep Neural Networks},
	year={2018},
	volume={},
	number={},
	pages={5089-5093},
}

@inproceedings{Zhang2019,
	author = {Zhang, Zixing and Wu, Bingwen and Schuller, Bj{\"{o}}rn},
	booktitle = {Proc. of 2019 IEEE Int. Conf. Acoust. Speech Signal Process. (ICASSP '19)},
	month = {May},
	pages = {6705--6709},
	title = {{Attention-augmented End-to-end Multi-task Learning for Emotion Prediction from Speech}},
	year = {2019}
}

@inproceedings{Duchi2010,
	author = {Duchi, John and Hazan, Elad and Singer, Yoram},
	booktitle = {23rd Conf. Learn. Theory (COLT 2010)},
	isbn = {9780982252925},
	pages = {257--269},
	title = {{Adaptive subgradient methods for online learning and stochastic optimization}},
	volume = {12},
	year = {2010}
}

@article{Hinton2015,
	title={Distilling the knowledge in a neural network},
	author={Hinton, Geoffrey and Vinyals, Oriol and Dean, Jeff},
	journal={arXiv:1503.02531},
	year={2015},
	url = {https://arxiv.org/abs/1503.02531},
}

@inproceedings{Latif2018,
	archivePrefix = {arXiv},
	arxivId = {1801.06353},
	author = {Latif, Siddique and Rana, Rajib and Younis, Shahzad and Qadir, Junaid and Epps, Julien},
	booktitle = {Proc. Annu. Conf. Int. Speech Commun. Assoc. (INTERSPEECH 2018)},
	doi = {10.21437/Interspeech.2018-1625},
	eprint = {1801.06353},
	issn = {19909772},
	pages = {257--261},
	title = {{Transfer learning for improving speech emotion classification accuracy}},
	year = {2018},
	month = {Sep.},
}

@inproceedings{Neumann2018,
	archivePrefix = {arXiv},
	arxivId = {1803.00357},
	author = {Neumann, Michael and {Thang Vu}, N. Goc},
	booktitle = {Proc. of 2018 IEEE Int. Conf. Acoust. Speech Signal Process. (ICASSP '18)},
	doi = {10.1109/ICASSP.2018.8462162},
	eprint = {1803.00357},
	isbn = {9781538646588},
	issn = {15206149},
	month = {Sep.},
	pages = {5769--5773},
	title = {{CRoss-lingual and Multilingual Speech Emotion Recognition on English and French}},
	year = {2018}
}

@article{Arjovsky2017,
	archivePrefix = {arXiv},
	journal = {arXiv:1701.07875},
	doi= {10.48550/arxiv.1701.07875},
	author = {Arjovsky, Martin and Chintala, Soumith and Bottou, L{\'{e}}on},
	title = {{Wasserstein GAN}},
	year = {2017}
}

@inproceedings{Schneider2019,
	author = {Schneider, Steffen and Baevski, Alexei and Collobert, Ronan and Auli, Michael},
	booktitle = {Proc. of Annu. Conf. Int. Speech Commun. Assoc. (INTERSPEECH 2019)},
	doi = {10.21437/Interspeech.2019-1873},
	issn = {19909772},
	pages = {3465--3469},
	title = {{wav2vec: Unsupervised pre-training for speech recognition}},
	month = {Sep.},
	year = {2019}
}

@article{Babu2021,
	author = {Babu, Arun and Wang, Changhan and Tjandra, Andros and Lakhotia, Kushal and Xu, Qiantong and Goyal, Naman and Singh, Kritika and von Platen, Patrick and Saraf, Yatharth and Pino, Juan and Baevski, Alexei and Conneau, Alexis and Auli, Michael},
	title = {{XLS-R: Self-supervised Cross-lingual Speech Representation Learning at Scale}},
	url = {http://arxiv.org/abs/2111.09296},
	year = {2021},
	journal = {arXiv:2111.09296}
}

@article{Livingstone2018,
	doi = {10.1371/journal.pone.0196391},
	author = {Livingstone, Steven R. AND Russo, Frank A.},
	journal = {PLOS ONE},
	publisher = {Public Library of Science},
	title = {The Ryerson Audio-Visual Database of Emotional Speech and Song (RAVDESS): A dynamic, multimodal set of facial and vocal expressions in North American English},
	year = {2018},
	month = {05},
	volume = {13},
	url = {https://doi.org/10.1371/journal.pone.0196391},
	pages = {1-35},
	number = {5},
}

@inproceedings{Schuller2004,
	author = {Schuller, Bj{\"{o}}rn and Rigol, Gerhard and Lang, Manfred},
	booktitle = {Proc. of 2004 IEEE Int. Conf. Acoust. Speech Signal Process. (ICASSP '04)},
	doi = {10.1109/icassp.2004.1326051},
	issn = {15206149},
	title = {{Speech emotion recognition combining acoustic features and linguistic information in a hybrid support vector machine - Belief network architecture}},
	volume = {1},
	year = {2004}
}

@inproceedings{Xiao2017,
	author = {Xiao, Zhongzhe and Wu, Di and Zhang, Xiaojun and Tao, Zhi},
	booktitle = {Proc. of 2016 IEEE Int. Conf. Prog. Informatics Comput. (PIC 2016)},
	doi = {10.1109/PIC.2016.7949505},
	isbn = {9781509034833},
	month = {Jun.},
	pages = {253--257},
	title = {{Speech emotion recognition cross language families: Mandarin vs. Western Languages}},
	year = {2017}
}

@inproceedings{Eyben2009,
	author = {Eyben, Florian and W{\"{o}}llmer, Martin and Schuller, Bj{\"{o}}rn},
	doi = {10.1109/ACII.2009.5349350},
	isbn = {9781424447992},
	booktitle = {Proc. of 3rd Int. Conf. Affect. Comput. Intell. Interact. Work. (ACII 2009)},
	title = {{OpenEAR - Introducing the Munich open-source emotion and affect recognition toolkit}},
	year = {2009}
}

@article{ElAyadi2011,
	author = {{El Ayadi}, Moataz and Kamel, Mohamed S and Karray, Fakhri},
	doi = {10.1016/j.patcog.2010.09.020},
	issn = {00313203},
	journal = {Pattern Recognit.},
	number = {3},
	pages = {572--587},
	title = {{Survey on speech emotion recognition: Features, classification schemes, and databases}},
	url = {www.elsevier.com/locate/pr},
	volume = {44},
	year = {2011}
}

@inproceedings{Li2021,
	author = {Li, Mao and Yang, Bo and Levy, Joshua and Stolcke, Andreas and Rozgic, Viktor and Matsoukas, Spyros and Papayiannis, Constantinos and Bone, Daniel and Wang, Chao},
	booktitle = {Proc. of 2021 IEEE Int. Conf. Acoust. Speech Signal Process. (ICASSP '21)},
	doi = {10.1109/ICASSP39728.2021.9413910},
	month = {May.},
	pages = {6329--6333},
	title = {{Contrastive Unsupervised Learning for Speech Emotion Recognition}},
	year = {2021}
}

@article{Zhang2020,
	author = {Zhang, Weijian and Song, Peng},
	doi = {10.1109/TASLP.2019.2955252},
	issn = {23299304},
	journal = {IEEE/ACM Trans. Audio Speech Lang. Process.},
	pages = {307--318},
	publisher = {Institute of Electrical and Electronics Engineers Inc.},
	title = {{Transfer sparse discriminant subspace learning for cross-corpus speech emotion recognition}},
	volume = {28},
	year = {2020}
}

@article{Eghbal-zadeh2018,
	archivePrefix = {arXiv},
	author = {Eghbal-zadeh, Hamid and Dorfer, Matthias and Widmer, Gerhard},
	title = {{Deep Within-Class Covariance Analysis for Robust Audio Representation Learning}},
	year = {2017},
	journal = {arXiv:1711.04022}
}

@inproceedings{Liang2019,
	author = {Liang, Jingjun and Chen, Shizhe and Zhao, Jinming and Jin, Qin and Liu, Haibo and Lu, Li},
	booktitle = {Proc. of 2019 IEEE Int. Conf. Acoust. Speech Signal Process. (ICASSP '19)},
	doi = {10.1109/ICASSP.2019.8683725},
	isbn = {9781479981311},
	issn = {15206149},
	month = {May.},
	pages = {4000--4004},
	title = {{Cross-culture Multimodal Emotion Recognition with Adversarial Learning}},
	year = {2019}
}

@inproceedings{Burkhardt2005,
	title={A database of German emotional speech.},
	author={Burkhardt, Felix and Paeschke, Astrid and Rolfes, Miriam and Sendlmeier, Walter F and Weiss, Benjamin and others},
	booktitle={Proc. Annu. Conf. Int. Speech Commun. Assoc. (INTERSPEECH 2005)},
	volume={5},
	pages={1517--1520},
	year={2005}
}

@article{Song2019,
	author = {Song, Peng},
	doi = {10.1109/TAFFC.2017.2705696},
	issn = {19493045},
	journal = {IEEE Trans. Affect. Comput.},
	month = {Apr.},
	number = {2},
	pages = {265--275},
	publisher = {Institute of Electrical and Electronics Engineers Inc.},
	title = {{Transfer linear subspace learning for cross-corpus speech emotion recognition}},
	volume = {10},
	year = {2019}
}

@inproceedings{Bergstra2013,
	author = {Bergstra, J and Yamins, D and Cox, D D},
	booktitle = {30th Int. Conf. Mach. Learn. ICML 2013},
	number = {PART 1},
	pages = {115--123},
	title = {{Making a science of model search: Hyperparameter optimization in hundreds of dimensions for vision architectures}},
	volume = {28},
	year = {2013}
}

@article{VandenOordDeepMinda,
	archivePrefix = {arXiv},
	author = {van den Oord, Aaron and Li, Yazhe and Vinyals, Oriol},
	title = {{Representation Learning with Contrastive Predictive Coding}},
	year = {2018},
	journal = {arXiv:1807.03748}
}

@article{Tang2021,
	author = {Tang, Duowei and Kuppens, Peter and Geurts, Luc and van Waterschoot, Toon},
	doi = {10.1186/s13636-021-00208-5},
	issn = {16874722},
	journal = {Eurasip J. Audio, Speech, Music Process.},
	month = {Dec.},
	number = {1},
	pages = {1--16},
	publisher = {Springer Science and Business Media Deutschland GmbH},
	title = {{End-to-end speech emotion recognition using a novel context-stacking dilated convolution neural network}},
	url = {https://asmp-eurasipjournals.springeropen.com/articles/10.1186/s13636-021-00208-5},
	volume = {2021},
	year = {2021}
}

@article{Gerczuk2021,
	author = {Gerczuk, Maurice and Amiriparian, Shahin and Ottl, Sandra and Schuller, Bj{\"{o}}rn},
	title = {{EmoNet: A Transfer Learning Framework for Multi-Corpus Speech Emotion Recognition}},
	year = {2021}, 
	journal = {IEEE Trans. Affect. Comput.}
}

@inproceedings{Rehman2020,
	author = {Rehman, Abdul and Liu, Zhen Tao and Li, Dan Yun and Wu, Bao Han},
	booktitle = {Proc. of Chinese Control Conf. (CCC 2020)},
	doi = {10.23919/CCC50068.2020.9189368},
	isbn = {9789881563903},
	issn = {21612927},
	month = {Jul.},
	pages = {7464--7468},
	title = {{Cross-Corpus Speech Emotion Recognition Based on Hybrid Neural Networks}},
	year = {2020}
}

@article{France2000,
	author = {France, Daniel J. and Shiavi, Richard G.},
	doi = {10.1109/10.846676},
	issn = {00189294},
	journal = {IEEE Trans. Biomed. Eng.},
	number = {7},
	pages = {829--837},
	pmid = {10916253},
	publisher = {IEEE},
	title = {{Acoustical properties of speech as indicators of depression and suicidal risk}},
	volume = {47},
	year = {2000}
}

@article{Hozjan2003,
	author = {Hozjan, Vladimir and Ka{\v{c}}i{\v{c}}, Zdravko},
	doi = {10.1023/A:1023426522496},
	issn = {13812416},
	journal = {Int. J. Speech Technol.},
	number = {3},
	pages = {311--320},
	title = {{Context-independent multilingual emotion recognition from speech signals}},
	volume = {6},
	year = {2003}
}

@article{Conneau2020,
	archivePrefix = {arXiv},
	author = {Conneau, Alexis and Baevski, Alexei and Collobert, Ronan and Mohamed, Abdelrahman and Auli, Michael},
	title = {{Unsupervised Cross-lingual Representation Learning for Speech Recognition}},
	year = {2020},
	journal = {arXiv:2006.13979}
}

@inproceedings{Parry2019,
	author = {Parry, Jack and Palaz, Dimitri and Clarke, Georgia and Lecomte, Pauline and Mead, Rebecca and Berger, Michael and Hofer, Gregor},
	booktitle = {Proc. of Annu. Conf. Int. Speech Commun. Assoc. (INTERSPEECH 2019)},
	doi = {10.21437/Interspeech.2019-2753},
	issn = {19909772},
	pages = {1656--1660},
	title = {{Analysis of deep learning architectures for cross-corpus speech emotion recognition}},
	url = {http://dx.doi.org/10.21437/Interspeech.2019-2753},
	year = {2019},
	month = {Sep.}
}

@article{Hafen,
	author = {Hafen, Ryan P and Henry, Michael J},
	journal = {Multimed. Syst.},
	doi = {10.1007/s00530-012-0266-0},
	issn = {09424962},
	number = {6},
	pages = {499--518},
	title = {{Speech information retrieval: A review}},
	volume = {18},
	year = {2012}
}

@article{Gideon2019,
	author = {Gideon, John and McInnis, Melvin G. and Provost, Emily Mower},
	journal = {IEEE Trans. Affect. Comput.},
	month = {Mar.},
	publisher = {arXiv},
	title = {{Improving cross-corpus speech emotion recognition with adversarial discriminative domain generalization ({ADDoG})}},
	year={2021},
	volume={12},
	number={4},
	pages={1055-1068},
	doi={10.1109/taffc.2019.2916092}
}

@inproceedings{Pepino2021,
	author = {Pepino, Leonardo and Riera, Pablo and Ferrer, Luciana},
	doi = {10.21437/interspeech.2021-703},
	pages = {3400--3404},
	title = {{Emotion Recognition from Speech Using wav2vec 2.0 Embeddings}},
	year = {2021},
	booktitle={Proc. of Annu. Conf. Int. Speech Commun. Assoc. (INTERSPEECH 2021)},
	month = {Sep.}
}

@inproceedings{sharma2022multi,
	title={Multi-lingual multi-task speech emotion recognition using wav2vec 2.0},
	author={Sharma, Mayank},
	booktitle={Proc. of 2022 IEEE Int. Conf. Acoust. Speech Signal Process. (ICASSP '21)},
	pages={6907--6911},
	year={2022},
}

@inproceedings{ma2024emotion2vec,
	title={emotion2vec: Self-supervised pre-training for speech emotion representation},
	author={Ma, Ziyang and Zheng, Zhisheng and Ye, Jiaxin and Li, Jinchao and Gao, Zhifu and Zhang, Shiliang and Chen, Xie},
	booktitle={Find. of the Assoc. for Comput. Linguist. ACL 2024},
	pages={15747--15760},
	year={2024}
}

@article{latif2022self,
	title={Self supervised adversarial domain adaptation for cross-corpus and cross-language speech emotion recognition},
	author={Latif, Siddique and Rana, Rajib and Khalifa, Sara and Jurdak, Raja and Schuller, Bj{\"o}rn},
	journal={IEEE Trans. Affect. Comput.},
	volume={14},
	number={3},
	pages={1912--1926},
	year={2022},
	publisher={IEEE}
}

@article{Krueger2014,
	author = {Alan B. Krueger  and Arthur A. Stone },
	title = {Progress in measuring subjective well-being},
	journal = {Science},
	volume = {346},
	number = {6205},
	pages = {42-43},
	year = {2014},
	doi = {10.1126/science.1256392},
}

@article{Saxena2020,
	author = {Saxena, Anvita and Khanna, Ashish and Gupta, Deepak},
	doi = {10.33969/ais.2020.21005},
	issn = {2642-2859},
	journal = {J. Artif. Intell. Syst.},
	number = {1},
	pages = {53--79},
	title = {{Emotion Recognition and Detection Methods: A Comprehensive Survey}},
	url = {http://creativecommons.org/licenses/by/4.0/},
	volume = {2},
	year = {2020}
}

@inproceedings{Zhou2021,
	archivePrefix = {arXiv},
	arxivId = {2010.14794},
	author = {Zhou, Kun and Sisman, Berrak and Liu, Rui and Li, Haizhou},
	booktitle = {Proc. of 2021 IEEE Int. Conf. Acoust. Speech Signal Process. (ICASSP '21)},
	doi = {10.1109/ICASSP39728.2021.9413391},
	eprint = {2010.14794},
	issn = {15206149},
	pages = {920--924},
	title = {{Seen and unseen emotional style transfer for voice conversion with a new emotional speech dataset}},
	month = {Jun.},
	year = {2021}
}

@inproceedings{Latif2019,
	author = {Latif, Siddique and Qadir, Junaid and Bilal, Muhammad},
	booktitle = {Proc. of 8th Int. Conf. Affect. Comput. Intell. Interact. (ACII 2019)},
	doi = {10.1109/ACII.2019.8925513},
	isbn = {9781728138886},
	month = {Sep.},
	title = {{Unsupervised Adversarial Domain Adaptation for Cross-Lingual Speech Emotion Recognition}},
	year = {2019}
}

@inproceedings{Hatch2006,
	author = {Hatch, Andrew O. and Stolcke, Andreas},
	booktitle = {Proc. of 2006 IEEE Int. Conf. Acoust. Speech Signal Process. (ICASSP '06)},
	doi = {10.1109/icassp.2006.1661343},
	isbn = {142440469X},
	issn = {15206149},
	title = {{Generalized linear kernels for one-versus-all classification: Application to speaker recognition}},
	volume = {5},
	year = {2006}
}

@article{Castillo2018,
	title={Emotion detection and regulation from personal assistant robot in smart environment},
	author={Castillo, Jos{\'e} Carlos and Castro-Gonz{\'a}lez, {\'A}lvaro and Alonso-Mart{\'\i}n, Fern{\'a}ndo and Fern{\'a}ndez-Caballero, Antonio and Salichs, Miguel {\'A}ngel},
	journal={Personal assistants: Emerging computational technologies},
	pages={179--195},
	year={2018},
	publisher={Springer},
}

@inproceedings{Bhaykar2013,
	author = {Bhaykar, Manav and Yadav, Jainath and Rao, K. Sreenivasa},
	booktitle = {2013 Natl. Conf. Commun. (NCC 2013)},
	doi = {10.1109/NCC.2013.6487998},
	isbn = {9781467359528},
	issn = {15503607},
	title = {{Speaker dependent, speaker independent and cross language emotion recognition from speech using GMM and HMM}},
	year = {2013}
}

@inproceedings{grill2020bootstrap,
	title={Bootstrap your own latent-a new approach to self-supervised learning},
	author={Grill, Jean-Bastien and Strub, Florian and Altch{\'e}, Florent and Tallec, Corentin and Richemond, Pierre and Buchatskaya, Elena and Doersch, Carl and Avila Pires, Bernardo and Guo, Zhaohan and Gheshlaghi Azar, Mohammad and others},
	journal={Proc. of IEEE Conf. Adv. Neural Inf. Process Syst. (NeurIPS 2020)},
	volume={33},
	pages={21271--21284},
	year={2020},
	month = {Dec.},
	address = {California, USA},
}

@inproceedings{Baevski2020,
	author = {Baevski, Alexei and Zhou, Henry and Mohamed, Abdelrahman and Auli, Michael},
	booktitle = {Proc. of IEEE Conf. Adv. Neural Inf. Process Syst. (NeurIPS 2020)},
	issn = {10495258},
	title = {{wav2vec 2.0: A framework for self-supervised learning of speech representations}},
	volume = {33},
	year = {2020},
	pages = {12449-12460},
	month = {Dec.},
	address = {California, USA},
}

\end{document}